\documentclass[12pt]{iopart}

\usepackage{iopams}  
\usepackage{color}
\usepackage[dvipsnames]{xcolor} 
\usepackage{pstricks,pst-fill,pst-text,pst-node,pstricks-add,tabularx}
\usepackage{tikz}
\usepackage{psfrag}
\usepackage{graphicx}
\usetikzlibrary{snakes}
\usetikzlibrary{patterns}
\usepackage{stackrel}
\usepackage{amsfonts}
\usepackage{bm}          
\usepackage{amssymb}

\newcommand{\be}{\begin{equation}}
\newcommand{\ee}{\end{equation}}
\newcommand{\ba}{\begin{eqnarray}}
\newcommand{\ea}{\end{eqnarray}}

\def\R{{\mathbb R}}

\def\N{{\mathbb N}}

\begin{document}

\title[Critical manifold of the kagome-lattice Potts model]{Critical manifold of the kagome-lattice Potts model}

\author{Jesper Lykke Jacobsen$^{1,2}$ and Christian R.\ Scullard$^{3}$}
\address{${}^1$LPTENS, \'Ecole Normale Sup\'erieure, 24 rue Lhomond, 75231 Paris, France}
\address{${}^2$Universit\'e Pierre et Marie Curie, 4 place Jussieu, 75252 Paris, France}
\address{${}^3$Lawrence Livermore National Laboratory, Livermore CA 94550, USA}

\eads{\mailto{jesper.jacobsen@ens.fr},
      \mailto{scullard1@llnl.gov}}

\begin{abstract}

  Any two-dimensional infinite regular lattice $G$ can be produced by
  tiling the plane with a finite subgraph $B \subseteq G$; we call $B$
  a basis of $G$.  We introduce a two-parameter graph polynomial
  $P_B(q,v)$ that depends on $B$ and its embedding in $G$.  The
  algebraic curve $P_B(q,v)=0$ is shown to provide an approximation to
  the critical manifold of the $q$-state Potts model, with coupling
  $v={\rm e}^K-1 $, defined on $G$.  This curve predicts the phase
  diagram not only in the physical ferromagnetic regime ($v>0$), but
  also in the antiferromagnetic ($v<0$) region, where analytical
  results are often difficult to obtain. For larger bases $B$ the
  approximations become increasingly accurate, and we conjecture that
  $P_B(q,v)=0$ provides the exact critical manifold in the limit of
  infinite $B$. Furthermore, for some lattices $G$---or for the Ising
  model ($q=2$) on any $G$---the polynomial $P_B(q,v)$ factorises for
  any choice of $B$: the zero set of the recurrent factor then
  provides the {\em exact} critical manifold. In this sense, the
  computation of $P_B(q,v)$ can be used to detect exact solvability of
  the Potts model on $G$.

  We illustrate the method for two choices of $G$: the square lattice,
  where the Potts model has been exactly solved, and the kagome
  lattice, where it has not. For the square lattice we correctly
  reproduce the known phase diagram, including the antiferromagnetic
  transition and the singularities in the Berker-Kadanoff phase at
  certain Beraha numbers. For the kagome lattice, taking the smallest
  basis with six edges we recover a well-known (but now refuted)
  conjecture of F.Y.~Wu. Larger bases provide successive improvements
  on this formula, giving a natural extension of Wu's approach.  We
  perform large-scale numerical computations for comparison and find
  excellent agreement with the polynomial predictions. For $v>0$ the
  accuracy of the predicted critical coupling $v_{\rm c}$ is of the
  order $10^{-4}$ or $10^{-5}$ for the 6-edge basis, and improves to
  $10^{-6}$ or $10^{-7}$ for the largest basis studied (with 36
  edges).

\end{abstract}

\noindent
{\em We dedicate this article to Professor Fa-Yueh Wu on the
  occasion of his 80th birthday.}

\section{Introduction}

The two-dimensional $q$-state Potts model \cite{Potts52} has been
extensively investigated over the last sixty years \cite{Wu82} and has
served as a testbed for the powerful tools of quantum integrability
\cite{Baxter_book}. It is defined by the reduced hamiltonian
\be
 -\beta {\cal H} = K \sum_{(ij) \in E} \delta(\sigma_i,\sigma_j) \,,
 \label{hamiltonian}
\ee
where $K$ is the dimensionless coupling between the spins $\sigma_i =
1,2,\ldots,q$ that are defined on the the vertices $i \in V$ of some
given connected graph $G=(V,E)$ with vertex set $V$ and edge
set $E$. Here $\beta = 1/k_{\rm B}T$, where $T$ is the temperature and
$k_{\rm B}$ the Boltzmann constant, and $\delta(\sigma_i,\sigma_j)$ is
Kronecker's delta function. From now on we suppose that $G$ is a piece
of a regular lattice, and we are primarily interested in the limit of
an infinite two-dimensional lattice.

The partition function $Z$ corresponding to (\ref{hamiltonian}) is
most conveniently expressed in the Fortuin-Kasteleyn representation
\cite{FK72}
\be
 Z = \sum_{\sigma} \exp(-\beta {\cal H})
   = \sum_{A \subseteq E} v^{|A|} q^{k(A)} \,,
 \label{FK_repr}
\ee
where $|A|$ denotes the number of edges in the subset $A$, and $k(A)$
is the number of connected components in the induced graph $G_A =
(V,A)$. The temperature parameter $v = {\rm e}^K - 1$ will be used
instead of $K$ in the following.  Note that the representation
(\ref{FK_repr}) makes sense for any $q \in \R$, although
(\ref{hamiltonian}) initially supposed that $q \in \N$. We shall also
admit any value of $v \in \R$, although only $v \ge -1$ corresponds to
a real coupling, $K \in \R$.

On a given lattice $G$, the Potts model can in general only be exactly
solved along certain curves in the $(q,v)$ plane. This is in sharp
contrast with the Ising model (alias the $q=2$ state Potts model),
which can be solved at any temperature $v$ \cite{Onsager44}. Another
major difference is that the Potts model has been solved only on a
very few lattices, including the square \cite{Baxter73} and triangular
lattices \cite{Baxter78} and certain decorated versions thereof
\cite{Wu10a}. On the contrary, the Ising model can be solved on
essentially any lattice (provided that suitable boundary conditions,
usually doubly periodic, are imposed).

A first step in the solution of the Potts model on $G$ is to determine
the critical manifold---or critical frontier---, by which we
understand the points in $(q,v)$ space at which the model stands at a
phase transition.%
\footnote{The exact solution on the square lattice \cite{Baxter73}
  shows that in the ferromagnetic regime $(v>0$) the nature of the
  phase transition is first order for $q>4$, and second order
  (continuous) for $0 \le q \le 4$.  The belief that this is true on
  any lattice is supported by other exact solutions \cite{Baxter78}
  and general field theoretical arguments. The nature of the
  transitions for $q<0$ and/or $v<0$ is obviously a more delicate
  question, which has been addressed only in the simplest cases
  \cite{JacSal06}.}
It is a remarkable fact that, at least in the solvable cases
\cite{Baxter73,Baxter82,Baxter78}, the loci of exact solvability
coincide precisely with the critical manifold.  It is equally
remarkable that the critical manifolds on the square
\cite{Baxter73,Baxter82}, triangular \cite{Baxter78} and hexagonal
(the dual of the triangular) lattices turn out to be given by simple
algebraic curves:
\ba
 (v^2-q)(v^2+4v+q) &=& 0 \,, \qquad \mbox{(square lattice)} 
 \label{sq_latt_cc} \\
 v^3 + 3v^2 - q    &=& 0 \,, \qquad \mbox{(triangular lattice)}
 \label{tri_latt_cc} \\
 v^3 - 3q v - q^2  &=& 0 \,. \qquad \mbox{(hexagonal lattice)}
 \label{hex_latt_cc}
\ea
It is an outstanding question of lattice statistics to determine
whether these features hold true more generally.

The case of the Potts model on the kagome lattice has attracted
particular attention. This is due not only to the practical
applications of this model, but also to the fact that the kagome
lattice is the ``simplest'' lattice on which an exact solution has not
yet been found. More than thirty years ago Wu conjectured \cite{Wu79}
that the critical manifold is given by the sixth-order algebraic curve
\be
 v^6 + 6 v^5 + 9 v^4 - 2 q v^3 - 12 q v^2 - 6 q^2 v - q^3 = 0 \,.
 \qquad \mbox{(kagome lattice)}
 \label{kagome_wu}
\ee
The derivation of (\ref{kagome_wu}) relied on a number of exact
equivalences and---crucially---a certain non-rigorous homogeneity
assumption.%
\footnote{For completeness we mention that Tsallis \cite{Tsallis82}
  has proposed an alternative conjecture for the kagome-lattice
  critical manifold.  This has however been shown definitely not to be
  the exact expression. Moreover it is less accurate than Wu's
  conjecture (\ref{kagome_wu}).}

The curve (\ref{kagome_wu}) has indeed a certain number of pleasing features:
\begin{enumerate}
\item For $q=2$ it factorises as $(v+1)^2 (v^4+4v^2-8v-8) = 0$. This
  situates the ferromagnetic phase transition at $v_{\rm c} =
  \sqrt{3+2\sqrt{3}}-1$, in agreement with the exact solution
  \cite{KanoNaya53}.%
  \footnote{The Ising model on the kagome and hexagonal lattices can
    be related by combining a star-triangle and a decoration-iteration
    transformation \cite{Barry88}. It follows that
    $(1+v_{\rm kag})^2 = 2(1+v_{\rm hex})-1$ in our notations. 
    Only one of the solutions to (\ref{hex_latt_cc}),
    $v_{\rm hex} = 1+\sqrt{3}$, leads to real solutions for
    $v_{\rm kag}$, namely $v_{\rm kag} = -1 \pm \sqrt{3 + 2 \sqrt{3}}$.}
\item It passes through the origin $(q,v)=(0,0)$ with infinite slope,
  giving as expected a model of spanning trees (see \cite{JSS05}).
\item It behaves asymptotically as $v^2 \propto q$ for $q \gg 1$,
  ensuring as expected first-order coexistence between the phases
  dominated by the terms $A=E$ and $A=\emptyset$ in (\ref{FK_repr}).
\end{enumerate}
Subsequent numerical results have however unambiguously established that
(\ref{kagome_wu}) is not an exact result:
\begin{enumerate}
\item For $q \to 1$ it places the bond percolation threshold at
  $p_{\rm c} = \frac{v_{\rm c}}{1+v_{\rm c}} = 0.524\,429\,717\cdots$, which is not correct but close enough that it took many years to be definitively ruled out \cite{ZiffSuding97}. Recent numerics \cite{Wu10b} gives $p_{\rm c} =
  0.524\,404\,978\,(5)$.
\item For $q = 3$ it gives $v_{\rm c} = 1.876\,269\,208\cdots$. This
  can be contrasted with the estimate \cite{Jensen97} $v_{\rm c} =
  1.876\,456\,(40)$, obtained by analysing the low-temperature series
  for the magnetisation, susceptibility, zero-field partition
  function, and specific heat. Numerical diagonalisation of the
  transfer matrix provides the more accurate estimate \cite{Wu10b}
  $v_{\rm c} = 1.876\,458\,(3)$.
\item For $q = 4$ the prediction of (\ref{kagome_wu}) is $v_{\rm c} =
  2.155\,842\,236\cdots$. This cannot be discriminated by the less
  precise series result \cite{Jensen97} $v_{\rm c} = 2.156\,1\,(5)$,
  but is ruled out by the transfer matrix result \cite{Wu10b} $v_{\rm
    c} = 2.156\,20\,(5)$.
\item Discrepancies have also been observed for large $q \gg 4$
  \cite{Monroe03}.
\end{enumerate}
Below we shall present further conclusive numerical evidence against
(\ref{kagome_wu}). Despite this numerical refutation of the
conjecture (\ref{kagome_wu}), it is nevertheless remarkable; viewed as
an approximation it is extraordinarily precise, the accuracy being of the
order $10^{-4}$ or $10^{-5}$ (see also \cite{Wu10b}).

In view of this evidence it is tempting to try to improve on
(\ref{kagome_wu}) by systematically fitting the numerics to other
low-order algebraic curves with reasonable (integer)
coefficients. Such attempts \cite{Jensen97,Salas08} have however
proved inconclusive.  Another line of research is to extend the
approximation (\ref{kagome_wu}) to other (decorated) lattices of the
kagome type \cite{Wu10a,Wu10b}.
%

In this paper we present a general method for obtaining approximations
to the critical manifold of the Potts models defined on any
two-dimensional regular lattice $G$. The infinite lattice $G$ is
obtained by tiling two-dimensional space by a certain finite subgraph
$B \subseteq G$ that we shall refer to as the {\em basis}. The
embedding of $B$ in $G$ determines exactly how $G$ is obtained as a
tiling by the motif $B$ and will be defined precisely
below. Corresponding to each embedded basis $B$ we define a
two-parameter graph polynomial $P_B(q,v)$ which is closely related to
the Tutte polynomial \cite{Sokal05}; the approximation to the critical
manifold then reads simply
\be
  P_B(q,v) = 0 \,.
  \label{PB_zero}
\ee
The precision of the approximation can be systematically improved by
increasing the size, in a sense to be made more precise in section
\ref{sec:discussion}, of $B$.

The definition of $P_B(q,v)$ conceals a number of remarkable features:
\begin{enumerate}
\item When not exact, the approximation (\ref{PB_zero}) turns out to
  be very precise, even for the smallest possible choice of $B$.
  The analogue of (\ref{kagome_wu}) can thus be computed by hand
  rather easily for any regular lattice of interest.
\item For some lattices $G$---or for the Ising model ($q=2$) on any
  $G$---the polynomial $P_B(q,v)$ factorises for any choice of
  $B$. The zero set of the recurrent factor then turns out to provide
  the {\em exact} critical manifold. In this sense, the computation of
  $P_B(q,v)$ can be used to detect whether the Potts model on $G$
  might be exactly solvable.
\item The determination of $P_B(q,v)$ for larger bases is well suited
  for exact computer-assisted calculations. We shall pursue this point
  of view in section~\ref{sec:general} below.
\end{enumerate}

Here we illustrate the general method and the above remarkable
features for the cases when $G$ is the square or the kagome lattice.

The square lattice serves as a benchmark, since its critical manifold
is known completely \cite{Baxter73,Baxter82,Saleur91,JacSal06}. The
result (\ref{sq_latt_cc}) is recovered by applying the general method
to the simplest choice where the basis $B$ consists of 4 edges. We
present extensions to larger bases consisting of 8, 16 and 32 edges that all contain the exact result
(\ref{sq_latt_cc}) as a factor; the remaining factor gives information
about the phase transitions at the Beraha numbers inside the
antiferromagnetic regime \cite{JacSal06}.

In the case of the kagome lattice, the simplest case when $B$ consists
of 6 edges reproduces Wu's conjecture (\ref{kagome_wu}). We then
present extensions to larger bases consisting of 12, 24 and 36 edges.
These extensions systematically improve the agreement of the predicted
critical manifold with the existing numerical results. In addition we
present improved numerical results for the critical manifold---both in
the ferromagnetic ($v>0$) and antiferromagnetic ($v<0$) regime---
obtained by exact diagonalisation of the transfer matrix.  Comparing
this to the result from the 36-edge basis we find an accuracy of the
order $10^{-6}$ or $10^{-7}$ in the ferromagnetic regime ($v>0$), and
both qualitative and quantitative improvements on (\ref{kagome_wu}) in
the antiferromagnetic regime ($v<0$).

The specialisation of these results to percolation (i.e., $P_B(1,v)$)
has previously been reported by one of us \cite{Scullard11}. Percolation polynomials for other lattices are reported in \cite{ScullardZiff08,ScullardZiff10,Scullard11JSM}.

The paper is organised as follows. We begin the next section by
reviewing the derivation of exact critical manifolds for
three-terminal triangular-type lattices. We then show how these
results may be used to define the graph polynomial $P_B(q,v)$ on any
regular lattice by the contraction-deletion algorithm and use this to
rederive Wu's conjectures \cite{Wu79} for the checkerboard and kagome
critical manifolds. In section \ref{sec:general} we describe how this
polynomial is calculated on larger bases using a computer program, and
use the following sections to report polynomials computed for the
square and kagome lattices on bases of up to 36 edges. In section
\ref{sec:numerics} we support our conjecture that the polynomial
approximations converge to the exact values by comparing our results
with numerical calculations of critical curves.

\section{The graph polynomial}

The triangular-lattice Potts model with arbitrary interactions in
up-pointing triangles is exactly solvable \cite{WuLin80}. In
particular, there exists a nice duality argument determining the
critical manifold exactly. Since this is the starting point of our
construction we begin by reviewing it.

\subsection{Triangular-lattice Potts model with interactions in
  up-pointing triangles}
\label{sec:Wu-Lin}

\begin{figure}
\begin{center}
\includegraphics{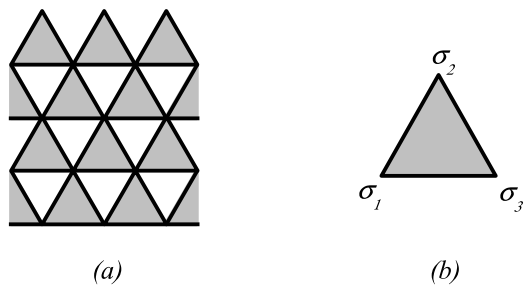}
\caption{a) Triangular lattice with arbitrary interactions in
  up-pointing triangles. b) Labelling of spins around an up-pointing
  triangle $\Delta_{123}$.}
\label{fig:triangular}
\end{center}
\end{figure}

Consider a Potts model on the triangular lattice in which all
interactions occur inside up-pointing triangles, as shown in
Fig.~\ref{fig:triangular}a. Let the three spins around an up-pointing
triangle $\Delta$ be labelled in cyclic order as $\sigma_1$,
$\sigma_2$, $\sigma_3$, starting from the spin in the lower-left
corner (Fig.~\ref{fig:triangular}b). The most general form of the
local Boltzmann weight on $\Delta_{123}$, compatible with the $S_Q$
permutational symmetry of the spins, takes the form
\be
 w_{123} = c_0 + c_1 \delta_{23} + c_2 \delta_{13} + c_3 \delta_{12} +
           c_4 \delta_{123} \,,
 \label{w123}
\ee
where we have introduced the short-hand notation $\delta_{ij} =
\delta(\sigma_i,\sigma_j)$ and $\delta_{ijk} =
\delta(\sigma_i,\sigma_j) \delta(\sigma_j,\sigma_k)$ for the Kronecker
delta functions.

Each term in the sum can be represented graphically as follows:
$$
\begin{pspicture}(-0.1,-0.4)(1.1,0.966)
 \psline[linewidth=0.5pt](0.0,0.0)(1.0,0.0)(0.5,0.866)(0.0,0.0)
 \pscircle*(0.0,0.0){0.1}
 \pscircle*(1.0,0.0){0.1}
 \pscircle*(0.5,0.866){0.1}
 \rput[Bc](0.5,-0.3){$c_0$}
\end{pspicture} \qquad
\begin{pspicture}(-0.1,-0.4)(1.1,0.966)
 \psline[linewidth=0.5pt](0.0,0.0)(1.0,0.0)(0.5,0.866)(0.0,0.0)
 \psline[linewidth=2.0pt,linecolor=blue](1.0,0.0)(0.5,0.289)(0.5,0.866)
 \pscircle*(0.0,0.0){0.1}
 \pscircle*(1.0,0.0){0.1}
 \pscircle*(0.5,0.866){0.1}
 \rput[Bc](0.5,-0.3){$c_1$}
\end{pspicture} \qquad
\begin{pspicture}(-0.1,-0.4)(1.1,0.966)
 \psline[linewidth=0.5pt](0.0,0.0)(1.0,0.0)(0.5,0.866)(0.0,0.0)
 \psline[linewidth=2.0pt,linecolor=blue](0.0,0.0)(0.5,0.289)(0.5,0.866)
 \pscircle*(0.0,0.0){0.1}
 \pscircle*(1.0,0.0){0.1}
 \pscircle*(0.5,0.866){0.1}
 \rput[Bc](0.5,-0.3){$c_2$}
\end{pspicture} \qquad
\begin{pspicture}(-0.1,-0.4)(1.1,0.966)
 \psline[linewidth=0.5pt](0.0,0.0)(1.0,0.0)(0.5,0.866)(0.0,0.0)
 \psline[linewidth=2.0pt,linecolor=blue](0.0,0.0)(0.5,0.289)(1.0,0.0)
 \pscircle*(0.0,0.0){0.1}
 \pscircle*(1.0,0.0){0.1}
 \pscircle*(0.5,0.866){0.1}
 \rput[Bc](0.5,-0.3){$c_3$}
\end{pspicture} \qquad
\begin{pspicture}(-0.1,-0.4)(1.1,0.966)
 \psline[linewidth=0.5pt](0.0,0.0)(1.0,0.0)(0.5,0.866)(0.0,0.0)
 \psline[linewidth=2.0pt,linecolor=blue](0.0,0.0)(0.5,0.289)(1.0,0.0)
 \psline[linewidth=2.0pt,linecolor=blue](0.5,0.289)(0.5,0.866)
 \pscircle*(0.0,0.0){0.1}
 \pscircle*(1.0,0.0){0.1}
 \pscircle*(0.5,0.866){0.1}
 \rput[Bc](0.5,-0.3){$c_4$}
\end{pspicture}
$$
Expand now the product over $\Delta_{123}$ to form the partition
function $Z$. This defines $Z$ in terms of clusters on the hexagonal
lattice $G = (V,E)$ with local weights $c_p$. Each cluster comes with
a weight $q$ from the summation over the spins $\sigma$, so that
\be
 Z = \sum_{A \subseteq E} q^{k(A)} \prod_{p=0}^4 (c_p)^{N_p} \,,
\ee
where $N_p$ is the number of up-triangles of type $c_p$, and $k(A)$ has
the same meaning as in (\ref{FK_repr}).

Alternatively, one can introduce loops that wrap around the boundaries
and internal cycles of the clusters. The loops are represented graphically
as follows:
$$
\begin{pspicture}(-0.1,-0.4)(1.1,0.966)
 \psline[linewidth=0.5pt](0.0,0.0)(1.0,0.0)(0.5,0.866)(0.0,0.0)
 \pscircle*(0.0,0.0){0.1}
 \pscircle*(1.0,0.0){0.1}
 \pscircle*(0.5,0.866){0.1}
 \psline[linearc=0.2,linecolor=red](-0.166,0.289)(0.5,0.289)(0.167,-0.289)
 \psline[linearc=0.2,linecolor=red](0.167,0.866)(0.5,0.289)(0.833,0.866)
 \psline[linearc=0.2,linecolor=red](1.166,0.289)(0.5,0.289)(0.833,-0.289)
 \rput[Bc](0.5,-0.3){$a_0$}
\end{pspicture} \qquad
\begin{pspicture}(-0.1,-0.4)(1.1,0.966)
 \psline[linewidth=0.5pt](0.0,0.0)(1.0,0.0)(0.5,0.866)(0.0,0.0)
 \pscircle*(0.0,0.0){0.1}
 \pscircle*(1.0,0.0){0.1}
 \pscircle*(0.5,0.866){0.1}
 \psline[linearc=0.2,linecolor=red](1.166,0.289)(0.5,0.289)(0.833,0.866)
 \psline[linearc=0.2,linecolor=red](0.167,0.866)(0.833,-0.289)
 \psline[linearc=0.2,linecolor=red](0.167,-0.289)(0.5,0.289)(-0.166,0.289)
 \rput[Bc](0.5,-0.3){$a_1$}
\end{pspicture} \qquad
\begin{pspicture}(-0.1,-0.4)(1.1,0.966)
 \psline[linewidth=0.5pt](0.0,0.0)(1.0,0.0)(0.5,0.866)(0.0,0.0)
 \pscircle*(0.0,0.0){0.1}
 \pscircle*(1.0,0.0){0.1}
 \pscircle*(0.5,0.866){0.1}
 \psline[linearc=0.2,linecolor=red](-0.166,0.289)(0.5,0.289)(0.167,0.866)
 \psline[linearc=0.2,linecolor=red](0.167,-0.289)(0.833,0.866)
 \psline[linearc=0.2,linecolor=red](0.833,-0.289)(0.5,0.289)(1.166,0.289)
 \rput[Bc](0.5,-0.3){$a_2$}
\end{pspicture} \qquad
\begin{pspicture}(-0.1,-0.4)(1.1,0.966)
 \psline[linewidth=0.5pt](0.0,0.0)(1.0,0.0)(0.5,0.866)(0.0,0.0)
 \pscircle*(0.0,0.0){0.1}
 \pscircle*(1.0,0.0){0.1}
 \pscircle*(0.5,0.866){0.1}
 \psline[linearc=0.2,linecolor=red](0.167,0.866)(0.5,0.289)(0.833,0.866)
 \psline[linearc=0.2,linecolor=red](-0.166,0.289)(1.166,0.289)
 \psline[linearc=0.2,linecolor=red](0.167,-0.289)(0.5,0.289)(0.833,-0.289)
 \rput[Bc](0.5,-0.3){$a_3$}
\end{pspicture} \qquad
\begin{pspicture}(-0.1,-0.4)(1.1,0.966)
 \psline[linewidth=0.5pt](0.0,0.0)(1.0,0.0)(0.5,0.866)(0.0,0.0)
 \pscircle*(0.0,0.0){0.1}
 \pscircle*(1.0,0.0){0.1}
 \pscircle*(0.5,0.866){0.1}
 \psline[linearc=0.2,linecolor=red](0.167,-0.289)(0.5,0.289)(0.833,-0.289)
 \psline[linearc=0.2,linecolor=red](0.833,0.866)(0.5,0.289)(1.166,0.289)
 \psline[linearc=0.2,linecolor=red](-0.166,0.289)(0.5,0.289)(0.167,0.866)
 \rput[Bc](0.5,-0.3){$a_4$}
\end{pspicture}
$$
and a local weight $a_p$ is associated with each diagram.  In
down-pointing triangles, there is no interaction and each face is of
the type $a_0$ (turned upside down).  One has therefore a model of
loops on a triangular lattice---shifted vertically by $1/\sqrt{3}$
lattice spacings with respect to the one on which the Potts spins are
defined---and each edge is covered by a loop.

Using the Euler relation the number of clusters $k(A)$ and the number
of loops $l(A)$ are related by $2k(A) = |V|-|A|+l(A)$, where we recall
that $A$ is the set of blue edges on the hexagonal lattice $G=(V,E)$.
The partition function in the loop representation therefore reads
\be
 Z = n^{|V|} \sum_{\rm loops} n^l \prod_{p=0}^4 (a_p)^{N_p} \,,
 \label{Wu-Lin}
\ee
where $n = \sqrt{q}$ is the weight of a loop, and the local weights
$a_p$ and $c_p$ are related by
\ba
   a_0 &=& c_0 \,, \\
 n a_i &=& c_i \quad \mbox{ for } i=1,2,3 \,, \\
 n^2 a_4 &=& c_4 \,.
 \label{a-type-weights}
\ea

Following \cite{WuLin80} we now suppose that at criticality the model
is invariant under a $\pi/3$ rotation of the triangular lattice where
the loops live.  This implies that the critical manifold is given by
$a_4 = a_0$ or \cite{WuLin80}
\be
 c_4 = q c_0 \,.
 \label{tri-duality}
\ee
Alternatively, the same relation can be found by a duality argument.
Indeed, if we first transform the triangular lattice into a hexagonal
lattice under duality, and next perform a decimation transformation of
one half of the spins, we recover the original triangular
lattice. Requiring that the weights be invariant under the combined
transformation again leads to (\ref{tri-duality}).

\subsection{Bases with three terminals}
\label{sec:3term}

As an example, we apply (\ref{tri-duality}) to a triangular lattice
with pure two-spin interactions, as in (\ref{FK_repr}).  We consider
the case of arbitrary inhomogeneous two-spin couplings
$\{v_1,v_2,v_3\}$ within $\Delta_{123}$, so that
\ba
 c_0 &=& 1 \,, \nonumber \\
 c_i &=& v_i \quad \mbox{ for } i=1,2,3 \,, \nonumber \\
 c_4 &=& v_1 v_2 v_3 + v_1 v_2 + v_2 v_3 + v_3 v_1 \,.
 \label{c4_tri}
\ea
The critical manifold is then given by (\ref{tri-duality}) in the form
(\ref{PB_zero}), where we have associated with the three-edge basis $B
= \Delta_{123}$ the graph polynomial
\be
  P_B(q,\{v_1,v_2,v_3\}) = v_1 v_2 v_3 + v_1 v_2 + v_2 v_3 + v_3 v_1 - q \,.
  \label{PB_tri3}
\ee
In the homogeneous case this indeed reduces to (\ref{tri_latt_cc}).

We can also recover the special case of the square lattice by setting
$v_3 = 0$ in (\ref{PB_tri3}). This formally corresponds to a two-edge
basis (remove one edge from $B$), and in the homogeneous case one
obtains the first factor in (\ref{sq_latt_cc}).

It should be noted that the result (\ref{tri-duality}) also applies
when extra spins are present inside the up-pointing triangles.%
\footnote{This has been used extensively in \cite{Wu10a}. One can also
  include explicit multi-spin interactions, but we shall not consider
  this possibility here.}
One then simply sums out those spins to get an effective interaction
of the form (\ref{w123}). As an example of this consider the hexagonal
lattice with arbitrary inhomogeneous two-spin couplings
$\{v_1,v_2,v_3\}$. We have then
\ba
 c_0 &=& v_1 + v_2 + v_3 + q \,, \nonumber \\
 c_1 &=& v_2 v_3 \,, \nonumber \\
 c_2 &=& v_3 v_1 \,, \nonumber \\
 c_3 &=& v_1 v_2 \,, \nonumber \\
 c_4 &=& v_1 v_2 v_3 \,.
 \label{c4_hex}
\ea
The critical manifold again results from (\ref{tri-duality}) in the form
(\ref{PB_zero}), with the graph polynomial
\be
  P_B(q,\{v_1,v_2,v_3\}) = v_1 v_2 v_3 - q(v_1 + v_2 + v_3 + q) \,.
  \label{PB_hex3}
\ee
In the homogeneous case we recover (\ref{hex_latt_cc}).

Note that (\ref{PB_tri3}) and (\ref{PB_hex3}) are related by the
duality transformation \cite{Wu82} $v_i v_i^* = q$, up to an
unimportant global factor. More precisely
\be
 P_{B^{\rm tri}}(q,\{v_1^*,v_2^*,v_3^*\}) =
 - \frac{q}{v_1 v_2 v_3} \, P_{B^{\rm hex}}(q,\{v_1,v_2,v_3\}) \,.
\ee

\subsection{Square-lattice Potts model with checkerboard interactions}
\label{sec:checkerboard}

Up to this point our determinations of the critical manifolds have
been exact.  We shall now see that crucial new ingredients appear when
we consider larger bases. In particular we pay attention to the way
the basis $B$ is embedded in $G$ so as to tile the entire lattice.

Let us call a vertex of $B$ a {\em terminal} if it will have to be
identified with a vertex of one or more copies of $B$ in the tiling of
$G$. The remaining vertices of $B$ are called {\em internal}.  The
models discussed in section~\ref{sec:Wu-Lin} used a basis with 3
terminals, and in the embedding of $B$ as up-pointing triangles each
terminal was glued to a terminal in two other copies of $B$. This
identification of each terminal with the other two is a convenient way
of characterising the embedding.

It should be clear from the examples (\ref{c4_tri}) and (\ref{c4_hex})
that the derivation of the weights $c_0$ and $c_4$ are closely
reminiscent of the computation of the partition function
(\ref{FK_repr}) itself. Indeed, these weights are just conditional
probabilities that the terminals are connected in a certain way (not
connected for $c_0$, and fully connected for $c_4$) in the
Fortuin-Kasteleyn expansion.

The partition function (\ref{FK_repr}) for the Potts model%
\footnote{In the mathematics literature, the Potts model partition function 
is known as the multivariate Tutte polynomial \cite{Sokal05}.}
defined on
any graph $G = (V,E)$ can be computed by the contraction-deletion
method (see \cite{Sokal05}). Namely, let $e \in E$ be any edge in
$G$. Denote by $G/e$ the graph obtained from $G$ by contracting $e$ to
a point and identifying the vertices at its end points (if they are
different); and denote by $G \setminus e$ the graph obtained from $G$
by deleting $e$. The terms in the sum (\ref{FK_repr}) can be grouped
in two disjoint classes, according to whether $e \in A$ or $e \notin
A$.  For the terms with $e \in A$ (resp.\ $e \notin A$), the number
$k(A)$ is unchanged upon contracting (resp.\ deleting) $e$. So the
obvious generalisation of (\ref{FK_repr}) to the case of arbitrary
edge-dependent weights satisfies
\be
 Z_G(q,\{v\}) = v_e Z_{G/e}(q,\{v\}) + Z_{G\setminus e}(q,\{v\}) \,.
 \label{cont_del}
\ee
Observing that $Z$ factorises over the components of a disconnected
graph, and that $Z=q$ for an isolated vertex, the contraction-deletion
formula (\ref{cont_del}) allows one to compute $Z_G(q,\{v\})$
recursively.

We now define the graph polynomial $P_B(q,\{v\})$ corresponding to
bases $B$ with more than three terminals. It is computed recursively
by applying the contraction-deletion formula (\ref{cont_del}) to
decrease the size of the basis. In the process we can rearrange $B$ by
identifying terminals from the embedding, and we shall avoid
disconnecting $B$ (modulo the embedding). The initial condition is to
replace any 3-terminal basis obtained in the recursion by its
corresponding critical manifold, which has been obtained in
section~\ref{sec:3term}.

Our key assumption---that we shall test in the following---is that
$P_B(q,\{v\}) = 0$ provides an approximation to the critical manifold
that becomes more and more precise (when it is not exact) as the size
of $B$ increases.

\begin{figure}
\begin{center}
\includegraphics{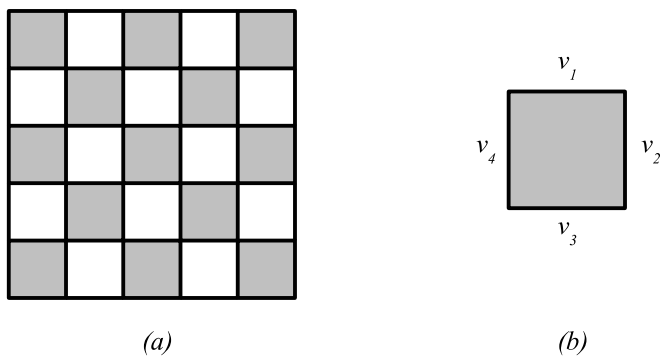}
\caption{a) The square lattice with the checkerboard couplings shown in b).}
\label{fig:checkerboard}
\end{center}
\end{figure}

Let us illustrate the above definition on a square-lattice Potts model
with interactions in alternating faces forming a checkerboard pattern;
see Fig.~\ref{fig:checkerboard}a. The corresponding basis has four
terminals, and the embedding amounts to gluing diametrically opposite
terminals. We are interested in the fully inhomogeneous case with
couplings $\{v_1,v_2,v_3,v_4\}$, as shown in
Fig.~\ref{fig:checkerboard}b.  Performing the deletion-contraction of
the edge with weight $v_4$ we obtain
\be
 P_B(q,\{v_1,v_2,v_3,v_4\}) = v_4 P_{B^{\rm tri}}(q,\{v_1,v_2,v_3\})
  + P_{B^{\rm hex}}(q,\{v_1,v_2,v_3\}) \,.
 \label{PB_tri_hex}
\ee
In the first term the two terminals have been identified, so we arrive
at the known case of the triangular lattice. In the second term we
have made use of the embedding to flip one of the edges.  In doing
that, one of the terminals is turned into an internal vertex, and we
recognise the 3-terminal basis of the hexagonal lattice. Substituting
now (\ref{PB_tri3}) and (\ref{PB_hex3}) into (\ref{PB_tri_hex}) we
find
\ba
 P_B(q,\{v_1,v_2,v_3,v_4\}) &=& v_1 v_2 v_3 v_4 +
 (v_2 v_3 v_4 + v_1 v_3 v_4 + v_1 v_2 v_4 + v_1 v_2 v_3) \nonumber \\
 & & -q (v_1 + v_2 + v_3 + v_4) - q^2 \,.
 \label{PB_check}
\ea
Note that this expression has an $S_4$ symmetry under the permutation
of any two couplings.

The expression $P_B(q,\{v\}) = 0$ with (\ref{PB_check}) was first
derived by Wu \cite{Wu79} using a different method (and a homogeneity
assumption). He initially conjectured that it was the exact critical
manifold.  However, Enting \cite{Enting87} has observed that the
putative $S_4$ symmetry is broken---albeit only at high order---in the
series expansions for various physical quantities of the
checkerboard-lattice Potts model, and he concluded that
(\ref{PB_check}) cannot be correct \cite{Wu_priv}.  In section 6.2 of
\cite{Maillard02} Maillard seems nevertheless to suggests that
(\ref{PB_check}) may be correct. Unfortunately we do not know of any
numerical investigation of the critical manifold for the completely
inhomogenious checkerboard Potts model that would allow us to assess
whether (\ref{PB_check}) is correct, and if not, how accurate an
approximation it is. But in section~\ref{sec:square} below we have
computed the polynomials $P_B(q,\{v\})$ corresponding to larger bases
for the square lattice, with four different couplings arranged in a
checkerboard pattern. We observe that these polynomials invariably
factorise, shedding the small factor (\ref{PB_check}). This is a
strong indication that (\ref{PB_check}) is indeed the exact critical
manifold of the checkerboard-lattice Potts model.

Something interesting happens if we specialise to the usual square
lattice with horizontal couplings $v_1=v_3$ and vertical couplings
$v_2 = v_4$. Then (\ref{PB_check}) factorises as
\be
 P_B(q,\{v_1,v_2,v_1,v_2\}) =
  (v_1 v_2-q)(v_1 v_2+2 v_1+2 v_2+q) \,,
 \label{PB_check_sq}
\ee
and in the homogeneous case we recover (\ref{sq_latt_cc}).  The zero
set of (\ref{PB_check_sq}) is known to be the exact critical
manifold. Indeed, the first factor corresponds to the ferromagnetic
transition curve \cite{Baxter73} and the second one to the
antiferromagnetic transition \cite{Baxter82} (see also
\cite{Saleur91,JacSal06}). It is particularly remarkable that the
antiferromagnetic transition curve is obtained exactly by this
approach, since---to our knowledge---it cannot be obtained by a simple
duality argument. Indeed, \cite{Baxter82} relies on the solution of
the Yang-Baxter equation and a so-called $Z$-invariance.

\subsection{Wu's conjecture recovered}

\begin{figure}
\begin{center}
\includegraphics{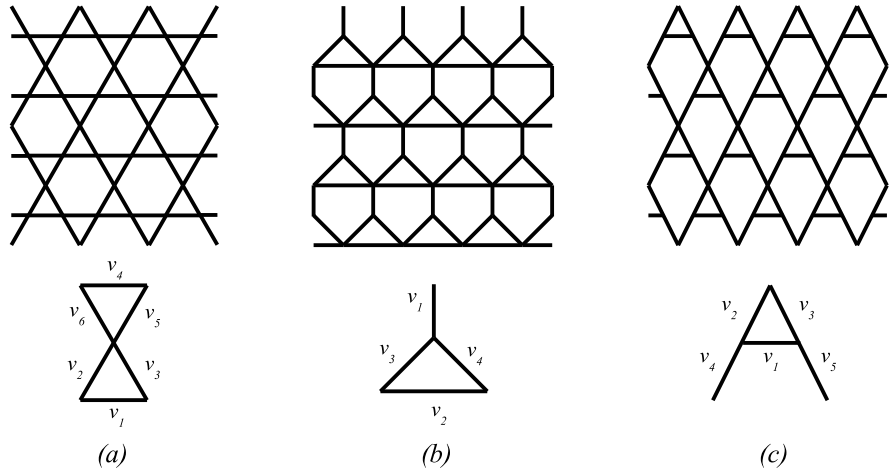}
\caption{Assignment of interactions on a) the kagome lattice; b) the martini-B lattice; c) the martini-A lattice.}
\label{fig:KAB}
\end{center}
\end{figure}

We now turn to the kagome lattice $G$. The simplest choice of basis
$B_6$ is the 4-terminal arrangement of six edges shown in
Fig.~\ref{fig:KAB}a which produces $G$ by the checkerboard embedding
discussed in section~\ref{sec:checkerboard}. Using the
contraction-deletion identity (\ref{cont_del}) on the $v_4$ edge we
obtain for the corresponding graph polynomial $K(\{v\}) =
P_{B_6}(q,\{v\})$:
\ba
 K(v_1,v_2,v_3,v_4,v_5,v_6) &=&
 v_4 B(v_5 + v_6 + v_5 v_6,v_1,v_2,v_3) \nonumber \\
 & & + A(v_1,v_2,v_3,v_5,v_6) \,.
 \label{KBA}
\ea
In the first term, the $v_4$ edge has been contracted, so that the
$v_5$ and the $v_6$ edges are in parallel. Using parallel reduction
\cite{Sokal05} these can then be replaced by a single edge with
coupling $v_5 + v_6 + v_5 v_6$. The result is a 3-terminal basis of the
martini-B lattice \cite{Scullard06,Ziff06} (see Fig.~\ref{fig:KAB}b) whose graph polynomial we have
denoted $B(v_1,v_2,v_3,v_4)$. This is then replaced by the critical
manifold (\ref{tri-duality}) which reads explicitly \cite{Wu06}
\ba
 B(v_1,v_2,v_3,v_4) &=& v_1 (v_2 v_3+v_2 v_4+v_3 v_4+v_2 v_3 v_4) \nonumber \\
 & &  -q(q+v_1+v_3+v_4) \,.
 \label{KB}
\ea
In the second term in (\ref{KBA}) the $v_4$ edge has been deleted.
Flipping the $v_5$ and $v_6$ edges using the embedding, the result is
a 3-terminal basis of the martini-A lattice (see Fig.~\ref{fig:KAB}c)
whose graph polynomial we have denoted $A(v_1,v_2,v_3,v_4,v_5)$. Using
again (\ref{tri-duality}) it is replaced by the critical manifold \cite{Wu06}
\ba
 A(v_1,v_2,v_3,v_4,v_5) &=& v_4 v_5(v_1 v_2+v_2 v_3+v_3 v_1+v_1 v_2 v_3)
 \nonumber \\
 & & -q(q^2 + q(v_1+v_2+v_3+v_4+v_5)+v_1 v_2 v_3
 \nonumber \\
 & & + (v_2+v_4)(v_3+v_5) + v_1(v_2+v_3+v_4+v_5)) \,.
 \label{KA}
\ea

Inserting (\ref{KB}) and (\ref{KA}) into (\ref{KBA}) produces the
desired graph polynomial. The corresponding approximation for the
critical manifold is $K(\{v\}) = 0$ by (\ref{PB_zero}). The case of a
kagome lattice with different couplings for each of the three
principal directions corresponds to the choice $v_4=v_1$, $v_5=v_2$
and $v_6=v_3$. This coincides with Wu's most general conjecture for
the critical manifold (see Eq.~(11) in \cite{Wu79}) and reduces to
(\ref{kagome_wu}) in the homogeneous case.

We have already discussed in the introduction that (\ref{kagome_wu})
is a very precise approximation to the critical manifold, but not an
exact result. Further numerical evidence for this statement will be
given in section~\ref{sec:numerics}.

\section{General bases} \label{sec:general}

As remarked above, the general contraction-deletion formula for
$P_B(q,\{v\})$ is identical to that for the partition function (see
(\ref{cont_del})), namely
\be
 P_B(q,\{v\}) = v_e P_{B/e}(q,\{v\}) + P_{B\setminus e}(q,\{v\}) \,. \label{eq:Pcont_del}
\ee
In the simple kagome example given above, both contraction and
deletion yielded lattices for which the exact solution is known. For
larger bases, this will not always be the case and then the algorithm
must be applied to whichever (or both) of $B/e$ and $B\setminus e$ is
unsolved, and so on recursively until solved lattices appear. This
results in a sort of binary tree in which each node is a lattice and
branches terminate on known graphs. However, the final polynomial is
independent of the edge chosen at each step, and is therefore a unique
property of the graph $B$ and the terminal identifications that define
its embedding in $G$.  Of course, the complexity of the problem
increases exponentially with the number of edges in the basis, and
thus the algorithm must be handled by computer for large bases. The
operation of the program that implements the contraction-deletion
algorithm can be sketched as follows:
\begin{enumerate}
\item As input, it takes an array of internal vertices, terminals and
  edges, along with the specification of their connectivities and
  identifications between terminals that determine how the basis is
  tiled to form the lattice.
\item An edge, $e$, is then chosen, respecting a limited set of
  criteria (e.g., not disconnecting the basis), for contraction and
  deletion to give the new bases $G/e$ and $G \setminus e$.
\item Often, these new graphs have edges doubled in series, or, as for
  the martini-B lattice in (\ref{KBA}), in parallel. These may be
  simplified by replacing them by a single edge with an effective
  interaction. Other complications are also possible, such as isolated
  or dead-end vertices, and these are removed from the problem.
\item After simplifying, $G/e$ and $G \setminus e$ are checked against
  a small set of known lattices. These are the square, hexagonal and
  triangular lattices, along with the one-dimensional chain and
  certain situations that are known to give critical manifolds that
  are identically zero (see \cite{Scullard12}). We restrict ourselves
  to these lattices because of the relative ease of identification
  (e.g., all edges are equivalent and we are thus not required to
  worry about edge-matching). If a graph is recognised, it is replaced
  by the corresponding graph polynomial and
  the process ends. If not, step 2 is called recursively with the
  unknown graph as input.
\item The final result is a tree of lattices with branches that
  terminate on known lattices. The output of the program is a list of
  functions giving the contraction-deletion formula for each
  lattice. This list is evaluated in {\sc Mathematica} and the
  homogeneous polynomial can then be found.
\end{enumerate}
This program was used to compute percolation ($q=1$) polynomials in
\cite{Scullard11} and \cite{Scullard12}. Only minor adaptations were
required to find the polynomials for general $q$, so we discuss only
those changes here and refer the interested reader to
\cite{Scullard12} where the implementation is described in detail.

\subsection{Simplification}

The only differences between the percolation and Potts implementations
are in the simplification phase of the algorithm, most importantly in
the way interactions are re-labelled upon the removal of doubled
bonds. The handling of dead-end vertices also requires a minor
modification.

\subsubsection{Edges doubled in parallel.}
Contracting the $v_1$ edge in the martini-A lattice of
Fig.~\ref{fig:KAB}c leaves $v_2$ and $v_3$ doubled in parallel. These
must be replaced by a single edge. In the percolation computation, the
probabilities $p_2$ and $p_3$ are combined into
$1-(1-p_2)(1-p_3)=p_2+p_3-p_2 p_3$, i.e., the probability that at
least one bond is open. The Potts expression for the effective
interaction, $v_p$, is similar \cite{Sokal05}:
\begin{equation}
 v_p = v_2+v_3+v_2 v_3 \ .
\end{equation}

\subsubsection{Edges doubled in series.}
For the martini-A lattice in Fig.~\ref{fig:KAB}c, deleting the $v_1$
edge leaves edges doubled in series, for example $v_3$ and
$v_5$. These are replaced by the effective interaction, $v_s$, given
by \cite{Sokal05}:
\be
 v_s=\frac{v_3 v_5}{v_3+v_5+q} \ . \label{eq:series}
\ee
The denominator in (\ref{eq:series}) presents a slight complication. Consider the deletion of $v_1$ in Fig.~\ref{fig:KAB}c, giving the square
lattice, which has the inhomogeneous manifold
$S(\tilde{v}_1,\tilde{v}_2)\equiv \tilde{v}_1 \tilde{v}_2 -q=0$, but with edges doubled in series. The
contraction-deletion formula (\ref{eq:Pcont_del}) works because the
completely inhomogeneous polynomial is at most first-order in any
interaction; we may have terms like $v_1 v_2 v_3$ but not $v_1^2 v_2$
or, more to the point, $v_1/(v_2+v_3+q)$. As such, we cannot use
\begin{equation}
 S\left(\frac{v_3 v_5}{v_3+v_5+q},\frac{v_2 v_4}{v_2+v_4+q}\right)
\end{equation}
for the manifold of $B\setminus e$. However, because the critical
manifold is found by setting $S(\tilde{v}_1,\tilde{v}_2)=0$, we are
free to multiply away these denominators, and we must do this before
inserting an expression into the contraction-deletion formula. Thus,
we have
\begin{eqnarray}
& &A(v_1,v_2,v_3,v_4,v_5)=v_1 H(v_2+v_3+v_2 v_3,v_4,v_5) \cr
&+&(v_3+v_5+q)(v_2+v_4+q)S\left(\frac{v_3 v_5}{v_3+v_5+q},\frac{v_2 v_4}{v_2+v_4+q}\right) ,
\end{eqnarray}
from which we recover (\ref{KA}).

\subsubsection{Dead-end vertices.}
\begin{figure}
\begin{center}
\includegraphics{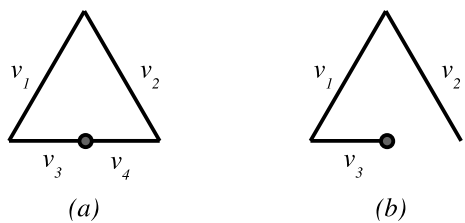}
\caption{a) Triangular lattice with a series-doubled edge; b) deletion
  of the $v_4$ edge leaves $v_3$ and its right vertex as a dead end.}
\label{fig:double_tri}
\end{center}
\end{figure}

An internal vertex connected to only one edge is a dead end. In
percolation, such an edge contributes nothing to the connectivities
between terminals, and therefore nothing to the graph polynomial. Its
probability, $p$, disappears from the problem because it only appears
in an overall factor of $[p+(1-p)]=1$ multiplying the probability of
every event. So the edge is simply removed from the problem along with
the vertex. In Potts language, the weight of every event is multiplied
by $(v+q)$ and thus, although we still remove the edge and vertex, the
manifold must be multiplied by this factor upon the removal. We give a
simple example. In Fig.~\ref{fig:double_tri}a, we have a triangular
lattice with two edges doubled in series. Adhering to the rules set
out in the previous section, the graph polynomial for this is given by
\begin{equation}
 (v_3+v_4+q)T\left(v_1,v_2,\frac{v_3 v_4}{v_3+v_4+q}\right) .
 \label{eq:double_tri}
\end{equation}
Contraction-deletion should also produce this answer. Contracting
$v_4$ just gives the simple triangular lattice, $T(v_1,v_2,v_3)$, but
deleting this edge results in the square lattice with a dead-end
vertex and edge, $v_3$. We remove these from the problem but retain
the factor of $(q+v_3)$ to give $(q+v_3)S(v_1,v_2)$. Injecting these
into the contraction-deletion formula,
\begin{equation}
 v_4 T(v_1,v_2,v_3)+(q+v_3)S(v_1,v_2),
\end{equation}
we do indeed recover (\ref{eq:double_tri}), which can be seen by
expanding both expressions.

Aside from these changes, the main operation of the program, including
edge selection and lattice identification, are exactly as described in
\cite{Scullard12}. We also have the same upper limit of practical feasibility,
namely using basis with at most 36 edges.

\section{Square-lattice Potts model using larger bases}
\label{sec:square}

Summarising, we have seen that the definition of $P_B(q,v)$ made in
section~\ref{sec:checkerboard} leads to known approximations of the
critical manifold when the lattice $G$ is obtained from a 4-terminal
basis.  In the remainder of the paper we shall corroborate our key
assumption that by using larger bases the accuracy of the
approximations can be improved, and more subtle features of the
critical manifolds can be uncovered.

To this end we study first the benchmark example of the square lattice
in this section. The following section will be devoted to the kagome
lattice.

\begin{figure}
\begin{center}
\includegraphics{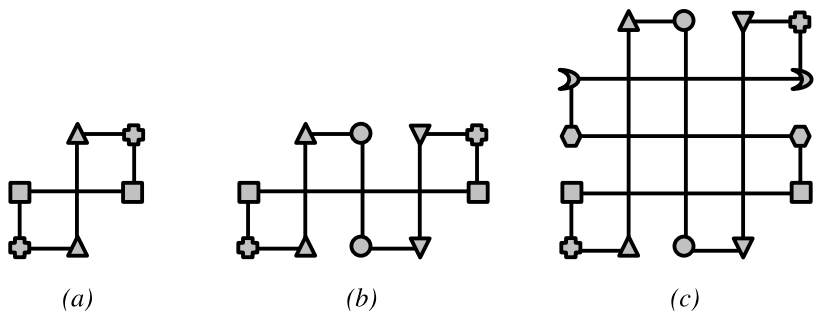}
\caption{a) 8-edge; b) 16-edge; and c) 32-edge bases for the square lattice. Terminals with matching shapes are identified.}
\label{fig:squarebases}
\end{center}
\end{figure}

We first extend the 4-edge checkerboard to the 8-edge basis in
Fig.~\ref{fig:squarebases}a. The polynomial for this case is given by
\begin{equation}
(q-v^2)(q + 4 v + v^2)(q^2+4qv+v^4+4 v^3+8 v^2) \,, \label{eq:square8}
\end{equation}
and we see the exact solutions appearing as factors. The polynomial
for the 16-edge basis depicted in Fig.~\ref{fig:squarebases}b is given
in the Appendix in equation (\ref{eq:square16}). We also found the
polynomial for the 32-edge basis in Fig.~\ref{fig:squarebases}c, which
is given in equation (\ref{eq:square32}).

We notice for the polynomials (\ref{eq:square8}), (\ref{eq:square16})
and (\ref{eq:square32}) that
\be
  P_B(q,v) = -v^{|E|} q^{-|E|/2} P_B(q,v^*)
\ee
with $v^* = q/v$, where $|E|$ is the number of edges in the basis. This
means that the graph polynomials respect the self-duality of the square
lattice.  In particular, the solutions to $P_B(q,v) = 0$ are either
self-dual curves, or pairs of mutually dual curves.

It appears that the exact solvability \cite{Baxter73,Baxter82} of the
square-lattice Potts model manifests itself in the presence of the
factors (\ref{sq_latt_cc}) in all the graph polynomials constructed
from larger bases, irrespective of their embedding. More generally, we
have found that for four different coupling constants arranged in a
checkerboard pattern (see Fig.~\ref{fig:checkerboard}), the
polynomials always contain the factor (\ref{PB_check}). It remains to
assess whether the remaining (in general non-factorisable) factor in
$P_B$ has any physical relevance.

\begin{figure}
\begin{center}
\includegraphics[width=0.7\textwidth]{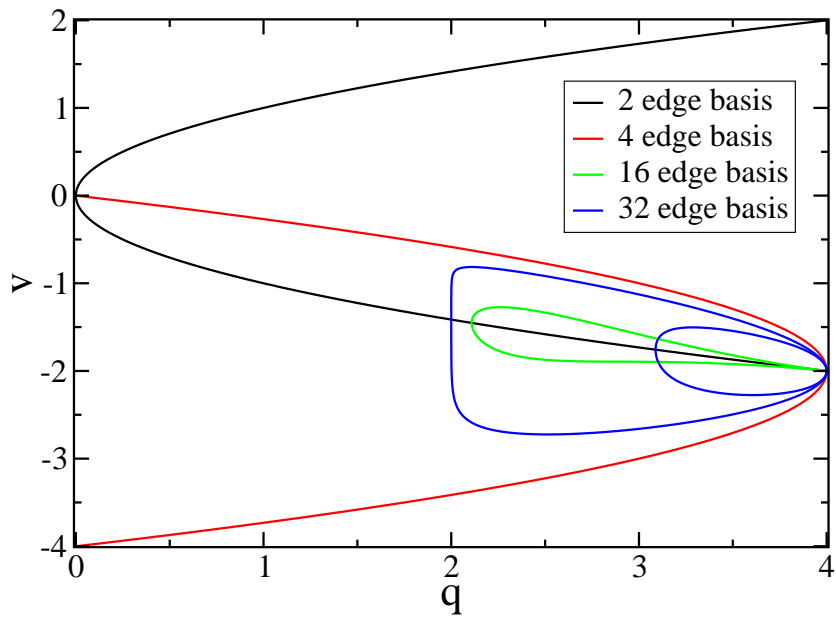}
\caption{Manifolds $P_B(q,v)$ for various choices of bases $B$ for the
  square lattice (see text). The exact critical manifolds
  (\ref{sq_latt_cc}) obtained from the bases with 2 and 4 edges are
  contained in the manifolds corresponding to larger bases.}
\label{fig:square}
\end{center}
\end{figure}

To investigate this issue, we have plotted the manifolds $P_B(q,v) =
0$ for the different bases in Fig.~\ref{fig:square}. It is seen that
the remaining factor produces dual pairs of curves inside the dual
pair of antiferromagnetic transition curves $v^2 + 4 v + q = 0$, in
the form of ``bubbles'' emanating from the point $(q,v)=(4,-2)$.  Each
bubble intersects the self-dual transition curve $v = -\sqrt{q}$ in two
points, of which one is exactly $q_0 = 4$ and the other we denote
$q_{\rm c}$. The bubbles are almost vertical in the vicinity of that
latter point.  For the 32-edge basis we find two bubbles with
respectively
\ba
 q_{{\rm c},1} &=& 2.000\,059\,024\cdots \,, \nonumber \\
 q_{{\rm c},2} &=& 3.088\,542\,134\cdots \,.
 \label{q12}
\ea
These values are conspicuously close to $2$ and $3$.

It was shown in \cite{Saleur91,JacSal06} that the Potts model
possesses singularities inside the so-called Berker-Kadanoff phase,
which is the region bordered by the antiferromagnetic transition
curves $v^2 + 4 v + q = 0$.  These singularities occur at the Beraha
numbers $q = B_k$ with
\be
 B_k = \left( 2 \cos (\pi/k) \right)^2 \,,
 \qquad \mbox{with $k=2,3,4,\ldots$}
 \label{Beraha}
\ee
and are {\em independent} of $v$. For a given integer $k$ the
singularity may or may not concern the {\em dominant} term in the
partition function, and only the former case leads to a phase
transition. The independence of $v$ implies that the incidence on the
phase diagram is the formation of a vertical ray in the $(q,v)$ plane.

The issue of dominance depends on the choice of boundary conditions.
In \cite{Salas06} it was argued from results of conformal field
theory---and checked numerically---that with cyclic boundary
conditions (free in one lattice direction and periodic in the other)
vertical rays are formed for $k \in 2 \N$. The corresponding result
for toroidal boundary conditions \cite{Salas07} (periodic in both
lattice direction) is that vertical rays occur only for $k=4$ and
$k=6$. Because of the identification of opposite terminals in the
embeddings of the basis $B$ that we have used, it is the toroidal
boundary conditions that are relevant in the present case.

In (\ref{q12}) we have indeed $q_{{\rm c},1} \approx B_4 = 2$ and
$q_{{\rm c},2} \approx B_6 = 3$, in agreement with \cite{Salas07}.  We
therefore conjecture that in the limit of an infinitely large basis,
$P_B(q,v)$ will contain (\ref{sq_latt_cc}) and a remaining factor
whose zeros produce a couple of vertical rays at $q=2$ and $q=3$,
extending between the two branches of the antiferromagnetic transition
curve $v^2 + 4 v + q = 0$.  The two bubbles obtained from the 32-edge
basis provide strong support of this conjecture; we believe that the
non-vertical parts of those bubbles will coincide with parts of the
antiferromagnetic transition curve in the limit of infinite $B$.

To produce an almost vertical ray from $P_B(q,v) = 0$ clearly requires
the algebraic expression $P_B(q,v)$ to be of large degree in $q$ and
$v$. Obviously the degree in $q$ equals the number of vertices in the
basis $B$ (counted up to identification of terminals through the
embedding), while the degree in $v$ equals the number of edges in
$B$. The above conjecture is compatible with the increase of degree as
the basis becomes larger: infinite-degree expressions are needed to
produce truly vertical rays in the limit of an infinite basis.

\section{Kagome-lattice Potts model using larger bases}
\label{sec:kagome}

\begin{figure}
\begin{center}
\includegraphics{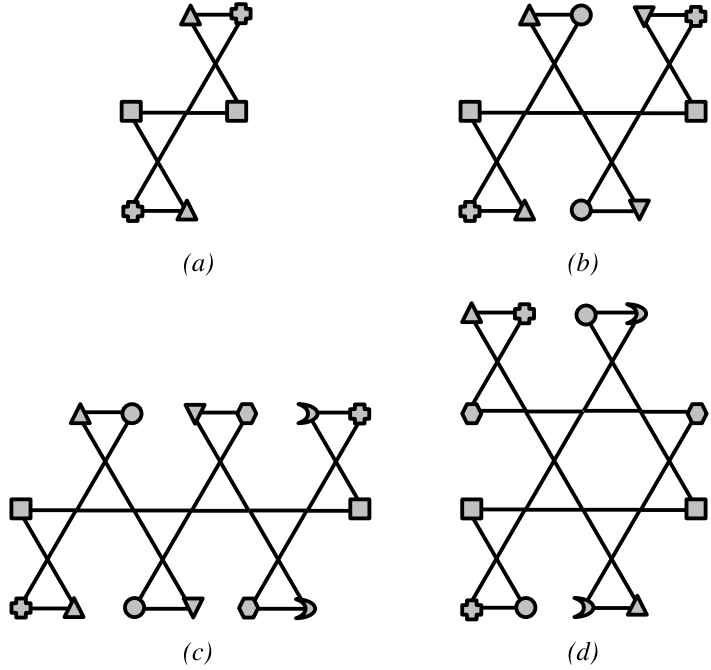}
\caption{a) 12-edge; b) 24-edge; and c), d) 36-edge bases for the kagome lattice.}
\label{fig:kagomebases}
\end{center}
\end{figure}

\begin{figure}
\begin{center}
\includegraphics[width=0.7\textwidth]{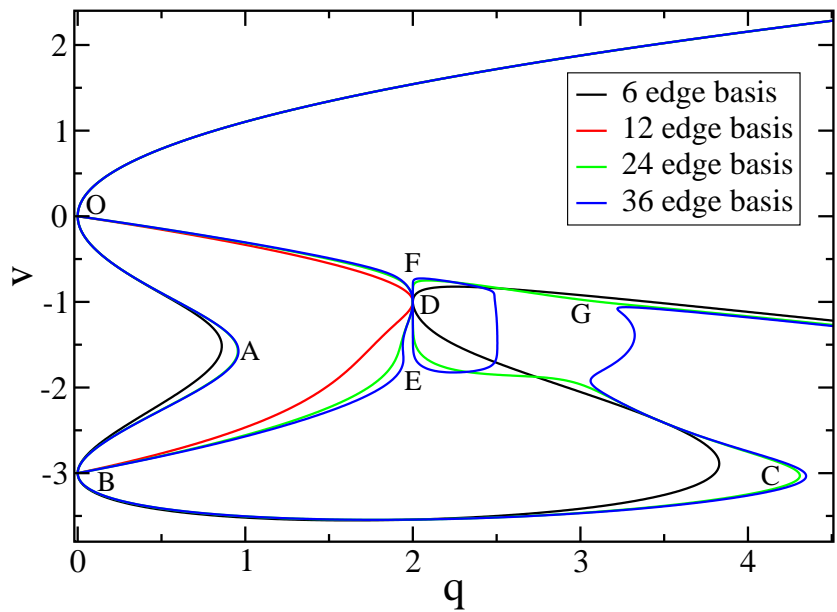}
\caption{Manifolds $P_B(q,v) = 0$ for various choices of bases $B$ for the
  kagome lattice (see text). The manifold corresponding to the basis
  with 6 edges is contained as a factor in the 12-edge manifold.
  The letters in the figure serve as a guide to the discussion in the main text.}
\label{fig:kagome}
\end{center}
\end{figure}

We are now ready to fulfill the main objective of this paper, which is
to improve on Wu's approximation (\ref{kagome_wu}) for the critical
manifold of the Potts model on the kagome lattice. Starting with the
$12$-edge basis in Fig.~\ref{fig:kagomebases}a, we find the polynomial
\begin{eqnarray}
& &(q^3 + 6 q^2 v + 12 q v^2 + 2 q v^3 - 9 v^4 - 6 v^5 - v^6) \times \cr
& &(q^3 + 6 q^2 v + 16 q v^2 + 24 v^3 + 2 q v^3 + 17 v^4 + 6 v^5 + v^6)=0 \ .
\label{12edge-kagome}
\end{eqnarray}
The first term in brackets is once again Wu's formula
(\ref{kagome_wu}), so, similar to the situation in percolation
\cite{Scullard11}, we get no better estimate of the ferromagnetic
critical point by extending from $6$ to $12$ edges. However, the
second term does provide us with additional information in the
antiferromagnetic region, which is plotted in red in
Fig.~\ref{fig:kagome}. We also found the polynomials for the 24-edge
basis of Fig.~\ref{fig:kagomebases}b, and the two different 36-edge
bases of Fig.~\ref{fig:kagomebases}c and \ref{fig:kagomebases}d. These
polynomials are given in the Appendix (see
eqs.~(\ref{eq:K4}), (\ref{eq:K2x3}), and (\ref{eq:K3x2})).

The manifolds $P_B(q,v) = 0$ for the different bases are plotted in
Fig.~\ref{fig:kagome}.

In the ferromagnetic region ($v>0$) the difference between the various
curves is not visible on the scale of the figure. The approximations for
the critical coupling in percolation read
\be
 v_{\rm c}(q=1) = \left \lbrace
 \begin{array}{lll}
 1.102\,738\,621\cdots & \mbox{(6-edge basis)} \\
 1.102\,636\,956\cdots & \mbox{(24-edge basis)} \\
 1.102\,632\,538\cdots & \mbox{(36-edge basis)} \\
 \end{array} \right.
\ee
to be compared with the most precise numerical estimate \cite{Wu10b}
$v_{\rm c}(q=1) = 1.102\,629\,24\,(2)$. The corresponding percolation
threshold is $p_{\rm c} = \frac{v_{\rm c}}{1+v_{\rm c}}$.
For the Ising model ($q=2$) all
curves pass through the exact critical point \cite{KanoNaya53} $v_{\rm
  c} = \sqrt{3 + 2\sqrt{3}}-1 = 1.542\,459\,756\cdots$. When $q=3$ we
have
\be
 v_{\rm c}(q=3) = \left \lbrace
 \begin{array}{lll}
 1.876\,269\,208\cdots & \mbox{(6-edge basis)} \\
 1.876\,439\,754\cdots & \mbox{(24-edge basis)} \\
 1.876\,447\,147\cdots & \mbox{(36-edge basis)} \\
 \end{array} \right.
\ee
to be compared with the numerical estimate \cite{Wu10b} $v_{\rm c}(q=3) =
1.876\,458\,(3)$. Finally, for $q=4$ we find
\be
 v_{\rm c}(q=4) = \left \lbrace
 \begin{array}{lll}
 2.155\,842\,236\cdots & \mbox{(6-edge basis)} \\
 2.156\,207\,452\cdots & \mbox{(24-edge basis)} \\
 2.156\,223\,187\cdots & \mbox{(36-edge basis)} \\
 \end{array} \right.
\ee
whereas the numerics \cite{Wu10b} gives $v_{\rm c}(q=4) = 2.156\,20\,(5)$. 

We defer a more detailed discussion of the ferromagnetic region to
section~\ref{sec:numerics}, where we shall also present our own
numerical results.  Instead we now discuss some common features and
differences of the curves and formulate some conjectures about the
thermodynamical limit.

There is a branch that extends to infinite $q$ in both the
ferromagnetic and antiferromagnetic ($v<0$) regions. Both of these
have the correct asymptotic behaviour $v \propto \sqrt{q}$ for all
bases.  Indeed, on a lattice of coordination number $z$, one needs $v
\propto q^{z/2}$ in order to ensure first-order phase coexistence
between the two dominant terms (with $A=E$ and $A=\emptyset$
respectively) in the expansion (\ref{FK_repr}).

With all bases, the ferromagnetic curve runs into the origin ${\rm O} :
(q,v)=(0,0)$ with infinite slope, as is required \cite{JSS05} to
produce a model of spanning trees. The continuation into the
antiferromagnetic region bends around at ${\rm A} : (q,v) \approx
(0.96,-1.58)$, and then goes down to the point ${\rm B} :(q,v)=(0,-3)$.  Since
all curves pass through $(q,v)=(0,-3)$ exactly, we conjecture that
this is a correct feature of the thermodynamical limit. If correct,
this would imply that on the dual (diced) lattice, the problem of
spanning forests \cite{JSS05} has a critical point with a
weight per tree $w_{\rm c} = -3$.

After bending around in ${\rm B}$ the curves continue in the region $v
< -3$ to a point ${\rm C} : (q,v) \approx (4.35,-3.03)$ where they
bend around again. It is interesting that ${\rm C}$ definitely has $q
> 4$, a feature that was not present in Wu's conjecture (alias the
6-edge curve).

The curves show strong finite-size effects in the domain $-3 < v < 0$
(excluding the branch ${\rm OAB}$ which has been discussed above).  In
particular, it might be that the point ${\rm C}$ will eventually move
to larger $q$, or even infinity.

For all bases the curves pass through ${\rm D} : (q,v) = (2,-1)$
exactly, and we conjecture that this is a correct feature of the
thermodynamical limit. For larger bases the branches ${\rm BD}$ and
${\rm DC}$ have a tendency to close up for $q \approx 2$ so as to
form an almost vertical ray. We conjecture that in the thermodynamical
limit there will be an exactly vertical ray extending between
${\rm E} : (q,v) = (2,v_1)$ and ${\rm F} : (q,v) = (2,v_2)$,
for some $v_1 < v_2$.%
\footnote{Note that the existence of this vertical ray is not in
  contradiction with the argument of Huse and Rutenberg \cite{Huse92}
  (using results of \cite{Barry88}) that the kagome-lattice Ising
  model is disordered for any $v > -1$. Indeed, the difference between
  sitting at $q=2$ exactly, and taking the limit $q \to 2$ in the
  $(q,v)$-plane phase diagram, is the quintessence of the special
  physics at the Beraha numbers within the antiferromagnetic region
  \cite{Saleur91,JacSal06}.}

It in then natural to assume that ${\rm BEC}$ and ${\rm OF}\infty$
will be smooth transition curves in the thermodynamical limit.
The 36-edge basis shows that the space in between these two curves is
likely to contain a further vertical ray. Since the properties of the
Berker-Kadanoff phase can be argued to be universal \cite{JacSal08} we
conjecture that the other ray will again be located at $q=3$
\cite{Salas07}, cf.\ our discussion of the square lattice above.  We further conjecture that the curves ${\rm OAB}$,
${\rm BEC}$ and ${\rm OF}\infty$ will surround the Berker-Kadanoff
phase. To delimit this phase to the right, one more curve is
needed. The 36-edge basis provides a convincing finite-size estimate
for this latter curve (starting at $(q,v) \approx (3.06,-1.92)$,
bending around at $(q,v) \approx (3.32,-1.39)$ and ending at $(q,v)
\approx (3.22,-1.08)$).

We believe that in the thermodynamical limit the curve $F\infty$ will
pass through the point ${\rm G} : (q,v) = (3,-1)$ exactly. Namely, one
can show \cite{MooreNewman00} that the three-state zero-temperature
antiferromagnet on the kagome lattice is equivalent to the
corresponding four-state model on the triangular lattice. The latter
is known to be critical with central charge $c=2$ (see
\cite{MooreNewman00} and references therein). Our finite bases locate
the antiferromagnetic transition in the $q=3$ model at
\be
 v_{\rm c}^{\rm AF}(q=3) = \left \lbrace
 \begin{array}{lll}
 -0.921\,400\,117\cdots & \mbox{(6-edge basis)} \\
 -0.973\,665\,377\cdots & \mbox{(24-edge basis)} \\
 \end{array} \right.
\ee
and it seems likely that this might tend to $v_{\rm c}^{\rm AF}(q=3) = -1$
in the thermodynamical limit. This is presumably the point in the phase diagram
where the finite-size effects are the most important.

We should stress here that to get a reliable picture of the phase
diagram it is not sufficient to study just one basis $B$, however
large. Indeed it is obvious from Fig.~\ref{fig:kagome} that different
bases reveal different parts of the critical manifold. For instance,
the 6-edge basis misses completely the curve ${\rm OFEB}$. And even
though the 36-edge basis gives in many respects the most precise
approximation to the true critical manifold, it misses a part of the
curve ${\rm F}\infty$ which is covered by the smaller 24-edge basis.

Quite obviously it would require more work to give numerical support
for the phase diagram discussed above and determine the relevant critical
properties. In section~\ref{sec:numerics} we limit ourselves to a
detailed investigation of the ferromagnetic transition curve and of
the antiferromagnetic branch ${\rm BC}$.

Note finally that the slope $1/w$ with which the curve ${\rm OD}$ goes
into the origin determines the critical point in a model of spanning
forests \cite{JSS05} on the kagome lattice. The critical weight per
component tree can be determined as
\be
 w_{\rm c} = \left \lbrace
 \begin{array}{ll}
 -3.364\,655\,607\cdots & \mbox{(12-edge basis)} \\
 -3.553\,344\,713\cdots & \mbox{(24-edge basis)} \\
 -3.578\,781\,346\cdots & \mbox{(36-edge basis)} \\
 \end{array} \right.
\ee

We can also determine the critical point of the flow polynomial
$\Phi(q)$, which is related to the Potts-model partition function
(\ref{FK_repr}) by setting $v=-q$ \cite{Sokal05}. We find
\be
 q_{\rm c} = \left \lbrace
 \begin{array}{ll}
 3.324\,717\,957\cdots & \mbox{(6-edge basis)} \\
 3.400\,923\,464\cdots & \mbox{(24-edge basis)} \\
 3.405\,701\,476\cdots & \mbox{(36-edge basis)} \\
 \end{array} \right.
 \label{qc-diced}
\ee
By duality this can also be interpreted as the critical point of the
chromatic polynomial on the diced lattice. Our results
(\ref{qc-diced}) agree with the crude extrapolation $q_{\rm c} \approx
3.4$ given in \cite{KSS08}.

\section{Numerical results}
\label{sec:numerics}

We have performed extensive numerical simulations in order to
accurately locate (parts of) the critical manifold for the
kagome-lattice Potts model. The objective is to assess whether the
graph polynomials found in section~\ref{sec:kagome} improve on Wu's
conjecture (\ref{kagome_wu})---and if so, by how much---when the size of
the basis increases.

For a critical system, the finite-size scaling of the free energy per
unit area $f(L)$ on a cylinder of circumference $L$ reads
\cite{cFSS1,cFSS2}
\be
 f(L) = f(\infty) - \frac{\pi c}{6 L^2} + o(L^{-2}) \,,
 \label{FSS}
\ee
where $f(\infty)$ is the free energy in the thermodynamical limit and
$c$ is the central charge. This result also provides an accurate means
of locating the critical point, since the numerically determined $c$
will exhibit a local extremum at criticality \cite{Zamolo}. The
precision can be improved \cite{Cardy98} by adding a non-universal
term $A/L^4$ to (\ref{FSS}).

\subsection{Transfer matrix construction}
\label{sec:TM}

The quantity $f(L)$ can be obtained from the largest eigenvalue of the
transfer matrix. There are two natural transfer directions for the
kagome lattice: parallel or perpendicular to one third of the lattice
edges. In Fig.~\ref{fig:kagomebases} the parallel and perpendicular
transfer directions are respectively horizontal and vertical.  Let
${\sf T}_N$ be the transfer matrix written in a basis of states that
keeps track of $N$ spins in a ``time slice''---a layer perpendicular
to the transfer direction. Using the basis provided by the
Fortuin-Kasteleyn representation (\ref{FK_repr}) then determines the
number of basis states (i.e., the dimension of ${\sf T}_N$) as the
Catalan number \cite{Blote82}
\be
 C_N = \frac{1}{N+1} {2N \choose N}
 \simeq \frac{4^N}{N^{3/2} \pi^{1/2}} \left[1 + {\cal O}(1/N)\right] \,.
\ee
The physical width $L$ is $N$ times the height (resp.\ $N$
times the side length) of an elementary triangle for the parallel
(resp.\ perpendicular) transfer direction. It is therefore most
efficient to choose the perpendicular direction (i.e., such that the
time runs vertically in Fig.~\ref{fig:kagomebases}). Setting the side
length of an elementary triangle to unity, we can henceforth identify
$N = L$.

We suppose henceforth $L$ even, since otherwise the kagome lattice
would not be compatible with periodic boundary conditions across the
time slice. Let ${\sf I}$, ${\sf J}_{i,i+1}$ and ${\sf D}_i$ be the
identity, join and detach operators (see \cite{SS01}) satisfying the
Temperley-Lieb algebra \cite{TL71} with weight $q$ per connected
component. The composite operators
\ba
 {\sf H}_i &=& {\sf I} + v {\sf J}_{i,i+1} \,, \nonumber \\
 {\sf V}_i &=& v {\sf I} + {\sf D}_i
\ea
then add respectively a horizontal (``space-like'') edge between the
vertices $i$ and $i+1$, and a vertical (``time-like'') edge a vertex
$i$. The edge weight $v$ is that of (\ref{FK_repr}). From these we can
define an operator that builds the basic bow tie pattern of the
kagome lattice:
\be
 {\sf B}_i = {\sf H}_i {\sf V}_i {\sf H}_i {\sf D}_{i+1}
             {\sf H}_i {\sf V}_i {\sf H}_i \,.
\ee
The transfer matrix then reads (by simple inspection of
Fig.~\ref{fig:kagomebases})
\be
 {\sf T}_L = \left( \prod_{i=1}^{L/2} {\sf B}_{2i} \right)
             \left( \prod_{i=1}^{L/2} {\sf B}_{2i-1} \right) \,.
\ee
Since this propagates the time slice by four times the height of an
elementary triangle, the corresponding free energy per unit area is
\be
 f(L) = - \frac{1}{2 \sqrt{3} L} \log \Lambda_{\rm max} \,,
\ee
where $\Lambda_{\rm max}$ is the eigenvalue of ${\sf T}_L$ with
maximal norm.

We have diagonalised ${\sf T}_L$ for $L=2,4,6,\ldots,16$ and computed
finite-size estimates $c(L)$ of the central charge $c$ from
(\ref{FSS}) with the $A/L^4$ term included, by using three successive sizes
$(L-4,L-2,L)$. For each fixed
$q=\frac{1}{10},\frac{2}{10},\ldots,\frac{40}{10}$ we have varied $v$
until a local extremum was found. This extremum provides an estimate
of the critical coupling.%
\footnote{For $q=1$ there is no finite-size dependence in $f(L)$, so
the method does not work in that case.} 

\subsection{Ferromagnetic critical curve}

We first present the results in the ferromagnetic region $v>0$.  It is
evident from Fig.~\ref{fig:kagome} that the various approximations to
the critical manifold are extremely close, so in order to appreciate
the differences we focus on the quantity $v-v_{24}$, where $v_{24}$ is
the relevant root of $P_B(q,v)$ using the 24-edge basis.

\begin{figure}
\begin{center}
\includegraphics[width=0.7\textwidth]{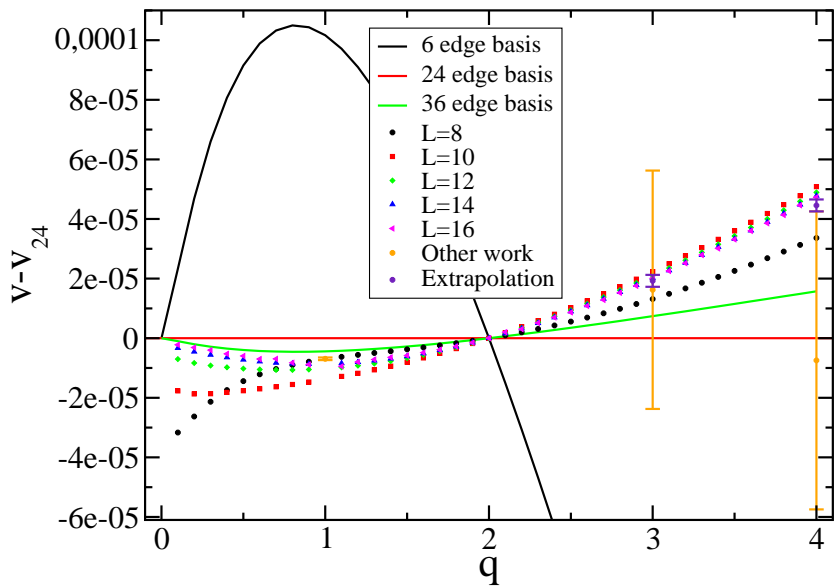}
\caption{Estimates of the critical curve in the ferromagnetic regime,
  relative to the approximation provided by the 24-edge basis. The
  solid lines show the zeros of $P_B(q,v)$ for various bases $B$. The
  points are numerical transfer matrix results for system size $L$
  (see text). Extrapolations for $L \to \infty$ are shown for $q=3$
  and $q=4$. Other numerical determinations are given
  for $q=1$, $q=3$, and $q=4$ (and labelled ``other work'').}
\label{fig:FM_diff24}
\end{center}
\end{figure}

Fig.~\ref{fig:FM_diff24} shows our numerical results (up to size
$L=16$) along with the approximations obtained from the graph
polynomials of section~\ref{sec:kagome} using bases with 6, 24 and 36
edges. For $q=3$ and $q=4$ we further show extrapolations of the
numerical results:
\be
 v_{\rm c} = \left \lbrace
 \begin{array}{ll}
 1.876\,459\,(2) & \qquad (q=3) \\
 2.156\,252\,(2) & \qquad (q=4) \\
 \end{array} \right.
 \label{vc_q34}
\ee
These were obtained by fitting the residual dependence in $1/L$ to first
and second-degree polynomials, and estimating the error bars by comparison
and successive elimination of the data points with small $L$.  We also show
in the figure the previous numerical results for $q=1$ (percolation
\cite{ZiffGu09,Wu10b}), $q=3$ (series expansions \cite{Jensen97}),
and $q=4$ (transfer matrix diagonalisations \cite{Wu10b}).%
\footnote{We do not show the series result \cite{Jensen97} for $q=4$,
  since the corresponding error bar is several times the vertical
  extent of Fig.~\ref{fig:FM_diff24}. The $q=3$ result of \cite{Wu10b}
  is also not shown, because it is almost identical to our own
  extrapolation.}

It is clear that at this level of precision the Wu conjecture
(\ref{kagome_wu})---i.e., the 6-edge basis result---is definitively
invalidated. More precisely, it is ruled out by our results
(\ref{vc_q34}) with a confidence level of $95$ (resp.\ $205$) standard
deviations for $q=3$ (resp. $q=4$).

More importantly, it is obvious from Fig.~\ref{fig:FM_diff24} that the
improved approximations with 24 and 36-edge bases move systematically
towards the numerical results. The results from the 36-edge base are
at a tiny distance from the numerics, of the order $10^{-6}$ or $10^{-7}$.
It is nevertheless clear that even those results only constitute an
approximation.

Note also that the numerics provides (within machine precision) the
exact $v_{\rm c}$ for the Ising model ($q=2$).

We finally remark that in a recent paper Baek {\em et al.}
\cite{Baek11} located the critical points for $q=3$ and $q=4$ using
phenomenological scaling of the internal energy crossings. These we
obtained from transfer matrices (with the ``parallel transfer
direction'' in the terminology of section~\ref{sec:TM}).  The authors
of \cite{Baek11} claimed that these results are independent of system
size, whence the results for the smallest possible systems ($N=2$ and
$N=4$ in the notation of section~\ref{sec:TM}):
\ba
 v_{\rm c} &=& 1.876\,313\,463\,895\cdots \,, \qquad \mbox{($q=3$)}
 \nonumber \\
 v_{\rm c} &=& 2.156\,174\,166\,284\cdots \,, \qquad \mbox{($q=4$)}
 \label{Baek}
\ea
would be exact. This claim is however not true.  Internal energy
crossings indeed exactly determine the critical point for the
square-lattice Potts model (by an easy duality argument), and for the
Ising model on more general lattices including kagome. However, for
the $q \neq 2$ Potts model on the kagome lattice, the energy crossings
{\em do} exhibit finite-size corrections. Note also that the results
(\ref{Baek}) are incompatible with our numerical results
(\ref{vc_q34}).

\subsection{Antiferromagnetic critical curve}

\begin{figure}
\begin{center}
\includegraphics[width=0.7\textwidth]{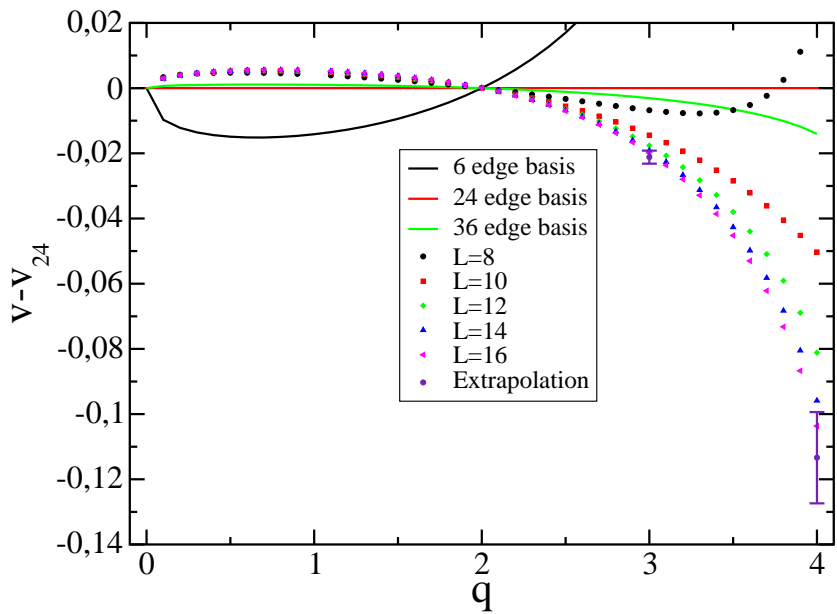}
\caption{Estimates of the critical curve ${\rm BC}$ in the
  antiferromagnetic regime, relative to the approximation provided by
  the 24-edge basis. The symbols and colours have the same meaning as
  in Fig.~\ref{fig:FM_diff24}.}
\label{fig:AF_diff24}
\end{center}
\end{figure}

We have similarly investigated numerically the antiferromagnetic
critical curve in the region $v < -3$ (referred to as ${\rm BC}$ in
section~\ref{sec:kagome}). As in the ferromagnetic case we show the
differences $v-v_{24}$ (see Fig.~\ref{fig:AF_diff24}).

It is evident from the scale of the vertical axis that the finite-size
effects are much larger than in the ferromagnetic region. In particular,
larger bases would be needed to convincingly approximate the numerical
results, especially in the region $q \simeq 4$.

One exception is again the Ising model ($q=2$) where finite-size
effects are completely absent. Indeed, all curves pass through the
exact value \cite{KanoNaya53} $v_{\rm c}(q=2) = -\sqrt{3+2\sqrt{3}}-1
= -3.542\,459\,756\cdots$.

Extrapolations of the finite-size numerics---obtained using the method
described above---read:
\be
 v_{\rm c} = \left \lbrace
 \begin{array}{ll}
 -3.486\,(2)  & \qquad (q=3) \\
 -3.361\,(14) & \qquad (q=4) \\
 \end{array} \right.
 \label{vcAF_q34}
\ee
and are shown in Fig.~\ref{fig:AF_diff24}. The result for $q=3$ is in
agreement with the Monte Carlo determination \cite{KSS08} of the
critical coupling of the 3-state Potts model on the diced lattice,
$v_{\rm c}^{\rm diced} = -0.860\,599\,(4)$. The corresponding dual value
$v_{\rm c} = 3 / v_{\rm c}^{\rm diced} = -3.485\,94\,(2)$ is in fact
more precise than (\ref{vcAF_q34}).

The Monte Carlo study \cite{KSS08} also shows that the 4-state Potts
model on the diced lattice is disordered for all $-1 \le v^{\rm diced}
< 0$. The corresponding dual statement is that for $q=4$, there is no
phase transition for $-\infty < v \le -4$. This again agrees with the
fact that our critical manifolds avoid this interval.

\subsection{Critical properties}

\begin{figure}
\begin{center}
\includegraphics[width=0.7\textwidth]{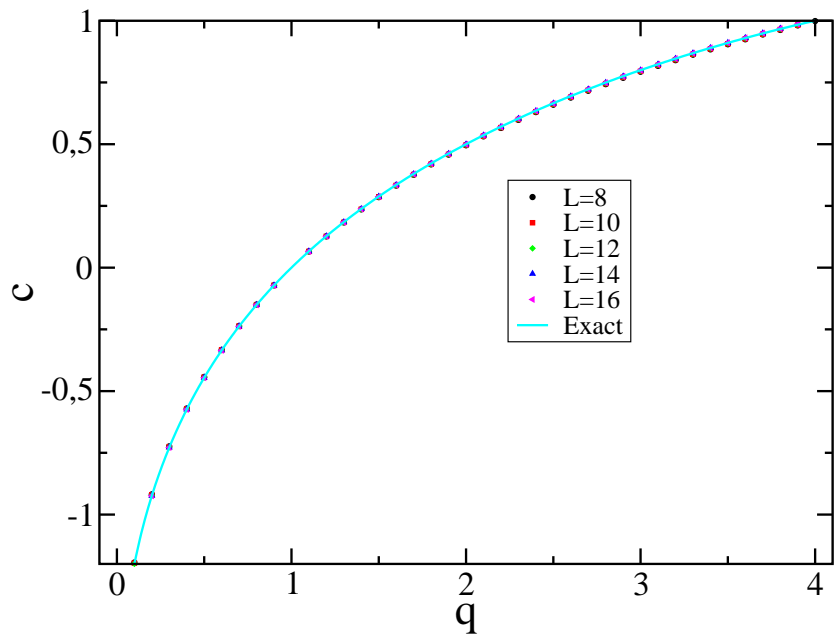}
\caption{Central charge $c$ along the ferromagnetic transition curve.}
\label{fig:cc_FM}
\end{center}
\end{figure}

The numerical results for the central charge along the ferromagnetic
transition curve are shown in Fig.~\ref{fig:cc_FM}. They agree with
the universality class of the usual ferromagnetic Potts model:
\be
 c = 1 - \frac{6}{k(k-1)} \,,
 \label{c_exact_FM}
\ee
where $q=B_k$ is parameterised by $k \in [2,\infty)$ through
  (\ref{Beraha}). In fact, the deviations from the exact result are
  not discernible on the scale of the figure.

\begin{figure}
\begin{center}
\includegraphics[width=0.7\textwidth]{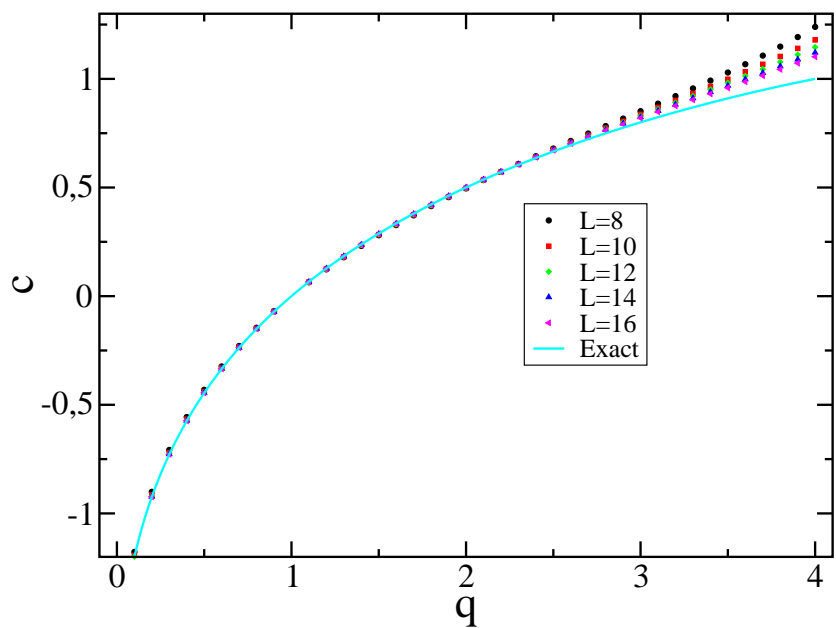}
\caption{Central charge $c$ along the antiferromagnetic transition
  curve ${\rm BC}$.}
\label{fig:cc_AF}
\end{center}
\end{figure}

The corresponding results along the antiferromagnetic curve ${\rm BC}$
are given in Fig.~\ref{fig:cc_AF}. Remarkably they again agree with
(\ref{c_exact_FM}). Note that there are now rather strong finite-size
effects in the region $q \simeq 4$.

The result for $k=6$---namely that the $q=3$ model at the
antiferromagnetic transition (\ref{vcAF_q34}) is in the universality
class of the three-state ferromagnetic model---was previously reported
in \cite{KSS08}.

\section{Discussion}
\label{sec:discussion}

In this article we have defined a graph polynomial $P_B(q,v)$ that
characterises a finite basis graph $B$ and the way it is embedded
in order to tile an infinite regular lattice $G$. Being closely
related to the Tutte polynomial, $P_B(q,v)$ can be evaluated
recursively from the contraction-deletion formula (\ref{cont_del}).

The zero sets (\ref{PB_zero}) were shown to reproduce well-known
approximations for the critical manifolds for the Potts model on the
checkerboard and kagome lattices \cite{Wu79} in the case
of the smallest possible bases. We have shown that such approximations
can be systematically ameliorated by increasing the size of the basis.
In particular, the 36-edge basis for the kagome lattice provides an
approximation of the critical manifold whose deviation from our
high-precision numerical study is of the order $10^{-6}$ or $10^{-7}$
in the ferromagnetic regime ($v>0$). In the antiferromagnetic regime
($v<0$) our approximations reveal qualitatively new structures, such
as singularities at the Beraha numbers (\ref{Beraha}) and the extent
of the Berker-Kadanoff phase \cite{Saleur91,JacSal06}.

The fact that the resulting 32nd-order algebraic curve still does not
reproduce perfectly the numerical data leads us to conjecture that the
{\em true} critical manifold for the kagome lattice is non-algebraic.
It is tempting to speculate that this behaviour may be responsible for
the fact that the Potts model on the kagome lattice has not yet been
solved (except for the $q=2$ Ising and $q=0$ spanning tree cases).

We should add here a small caveat. We have seen that both for
percolation \cite{Scullard12} and for the general $q$-state Potts
model, when they are not exact, the polynomials $P_B(q,v)$ provide
accurate approximations for unsolved problems that improve with an
increasing number of edges in $B$. This is the motivation for our
conjecture that the polynomials converge to the exact critical
manifold in the limit of certain infinite bases. However, as shown in
\cite{Scullard12}, it is not enough that the number of edges in $B$
tends to infinity.  Let us say that a generic basis is of size $N
\times M$ if it contains $N$ unit cells in the vertical direction and
$M$ unit cells in the horizontal direction. For example, for the
kagome lattice the unit cell is the six-edge bow tie pattern shown in
Fig.~\ref{fig:KAB}a, so that the bases of Figs.~\ref{fig:kagomebases}c
and \ref{fig:kagomebases}d are of size $2 \times 3$ and $3 \times 2$
respectively. In \cite{Scullard12}, an example was found of a lattice
for which the $1 \times M$ basis prediction for increasing $M$
converged to a number that, although similar to the numerically known
percolation threshold, was ruled out by simulations. Thus we should
add the condition that in order to produce a sequence of algebraic
curves of increasing order that converge to the exact critical
manifold, both $N$ and $M$ must go to infinity simultaneously with a
{\em finite} aspect ratio $\eta \equiv N/M \in (0,\infty)$.  Although
it is difficult to probe this at present, it seems at least plausible,
if not likely, that these infinite basis predictions should also be
independent of $\eta$.

An interesting feature of our findings is that in some sense the
computation of the polynomials $P_B(q,v)$ acts as a detector of exact
solvability for the underlying model. To be more precise, it seems
that if the polynomial factorises for any choice of basis, shedding
always the same ``small'' factor, the zero set of that factor provides
(a part of) the exact critical manifold. We have seen this mechanism
at play in the case of the square lattice, where the small factor is
given by (\ref{sq_latt_cc}). More generally, for the square lattice
with checkerboard interactions (see Fig.~\ref{fig:checkerboard}) the
small factor is given by (\ref{PB_check}), leading us to conjecture
that this is the exact critical manifold for the checkerboard model.
We stress that exact results are only expected if the factorisation is
systematic, i.e., occurs for any $B$. As an example of the contrary,
the 12-edge basis of the kagome lattice led to the fortuitous
factorisation (\ref{12edge-kagome}), and we have shown convincingly
that neither of the factors provide an exact result. We should also
notice that the implication does not seem to work the other way
around: exact solvability does not imply factorisation. This is
witnessed in particular by the case $(q,v)=(3,-1)$ of the
kagome-lattice Potts model, which is exactly solvable
\cite{MooreNewman00}, but not a zero of $P_B(q,v)$ for the finite
bases that we have studied here.

Our work hints at several directions for future research. It would
obviously be worthwhile having a more efficient means of dealing with
yet larger bases. This would require finding an alternative definition
of $P_B(q,v)$ that does not refer to contraction-deletion. The study
of the critical manifolds for other lattices is another possibility.
Finally, it is quite possible that $P_B(q,v)$ conceals some graph
theoretical applications.

\section*{Acknowledgements}

The work of JLJ was supported by the Agence Nationale de la Recherche
(grant ANR-10-BLAN-0414:~DIME) and the Institut Universitaire de
France. This work was partially (CRS) performed under the auspices of
the U.S. Department of Energy by Lawrence Livermore National
Laboratory under Contract DE-AC52-07NA27344. JLJ wishes to express his
gratitude to Jes\'us Salas for collaboration on a related project
\cite{Salas08}. Both authors thank J.\ Salas and F.Y.\ Wu for some interesting
comments on the manuscript.
We are also grateful to the Mathematical Sciences Research
Institute at the University of California, Berkeley for hospitality
during the programme on Random Spatial Processes where this work was
initiated.

\renewcommand{\theequation}{A\arabic{equation}}
\setcounter{equation}{0}  
\section*{Appendix}

Here, we report the graph polynomials $P_B(q,v)$ for the various bases
$B$ and graphs $G$ considered.

\subsection*{Square lattice}

\noindent
The 16-edge basis of Fig.~\ref{fig:squarebases}b:
\begin{eqnarray}
& &(q - v^2) (q + 4 v + v^2) (q^6 + 12 q^5 v + 68 q^4 v^2 + 232 q^3 v^3 + 4 q^4 v^3 \cr
&+& 516 q^2 v^4 + 40 q^3 v^4 + q^4 v^4 + 736 q v^5 + 192 q^2 v^5 + 8 q^3 v^5 + 576 v^6 \cr
&+& 504 q v^6 + 52 q^2 v^6 + 736 v^7 + 192 q v^7 + 8 q^2 v^7 + 516 v^8 + 40 q v^8 \cr
&+& q^2 v^8 + 232 v^9 + 4 q v^9 + 68 v^{10} + 12 v^{11} + v^{12}) \label{eq:square16}
\end{eqnarray}

\noindent
The 32-edge basis of Fig.~\ref{fig:squarebases}c:
\begin{eqnarray}
& &(q - v^2) (q + 4 v + v^2) (q^{14} + 28 q^{13} v + 384 q^{12} v^2 + 3424 q^{11} v^3 \cr
&+&4 q^{12} v^3 + 22240 q^{10} v^4 + 112 q^{11} v^4 +    q^{12} v^4 + 111744 q^9 v^5 \cr
&+& 1536 q^{10} v^5 + 24 q^{11} v^5 + 450016 q^8 v^6 + 13632 q^9 v^6 + 304 q^{10} v^6 \cr
&+& 1484032 q^7 v^7 + 87520 q^8 v^7 + 2688 q^9 v^7 + 8 q^{10} v^7 + 4053840 q^6 v^8 \cr
&+& 430560 q^7 v^8 + 18520 q^8 v^8 + 176 q^9 v^8 + q^{10} v^8 + 9198080 q^5 v^9 \cr
&+& 1676320 q^6 v^9 + 104032 q^7 v^9 + 2048 q^8 v^9 + 20 q^9 v^9 + 17211584 q^4 v^{10} \cr
&+& 5249504 q^5 v^{10} + 484160 q^6 v^{10} + 16288 q^7 v^{10} + 256 q^8 v^{10} \cr
&+& 25996800 q^3 v^{11} + 13255488 q^4 v^{11} + 1867840 q^5 v^{11} + 98816 q^6 v^{11} \cr
&+& 2368 q^7 v^{11} + 12 q^8 v^{11} + 30259968 q^2 v^{12} + 26606400 q^3 v^{12} \cr
&+& 5923440 q^4 v^{12} + 479248 q^5 v^{12} + 17256 q^6 v^{12} + 208 q^7 v^{12} + q^8 v^{12}\cr
&+& 24551424 q v^{13} + 40761344 q^2 v^{13} + 15114304 q^3 v^{13} + 1896032 q^4 v^{13} \cr
&+& 100736 q^5 v^{13} + 2240 q^6 v^{13} + 16 q^7 v^{13} + 10616832 v^{14} + 43298304 q v^{14} \cr
&+& 29706752 q^2 v^{14} + 6060928 q^3 v^{14} + 484352 q^4 v^{14} + 16800 q^5 v^{14} \cr
&+& 240 q^6 v^{14} + 24551424 v^{15} + 40761344 q v^{15} + 15114304 q^2 v^{15} \cr
&+& 1896032 q^3 v^{15} + 100736 q^4 v^{15} + 2240 q^5 v^{15} + 16 q^6 v^{15} + 30259968 v^{16}\cr
&+& 26606400 q v^{16} + 5923440 q^2 v^{16} + 479248 q^3 v^{16} + 17256 q^4 v^{16} \cr
&+& 208 q^5 v^{16} + q^6 v^{16} + 25996800 v^{17} + 13255488 q v^{17} + 1867840 q^2 v^{17}\cr
&+& 98816 q^3 v^{17} + 2368 q^4 v^{17} + 12 q^5 v^{17} + 17211584 v^{18} + 5249504 q v^{18}\cr
&+& 484160 q^2 v^{18} + 16288 q^3 v^{18} + 256 q^4 v^{18} + 9198080 v^{19} + 1676320 q v^{19}\cr
&+& 104032 q^2 v^{19} + 2048 q^3 v^{19} + 20 q^4 v^{19} + 4053840 v^{20} + 430560 q v^{20} \cr
&+& 18520 q^2 v^{20} + 176 q^3 v^{20} + q^4 v^{20} + 1484032 v^{21} + 87520 q v^{21} \cr
&+& 2688 q^2 v^{21} + 8 q^3 v^{21} + 450016 v^{22} + 13632 q v^{22} + 304 q^2 v^{22} \cr
&+& 111744 v^{23} + 1536 q v^{23} + 24 q^2 v^{23} + 22240 v^{24} + 112 q v^{24} + q^2 v^{24}\cr
&+& 3424 v^{25} + 4 q v^{25} + 384 v^{26} + 28 v^{27} + v^{28}) \label{eq:square32}
\end{eqnarray}

\subsection*{Kagome lattice}

\noindent
The 24-edge basis of Fig.~\ref{fig:kagomebases}b:
\begin{eqnarray}
& &q^{12} + 24 q^{11} v + 276 q^{10} v^2 + 2016 q^9 v^3 + 8 q^{10} v^3 + 10452 q^8 v^4 \cr
&+& 168 q^9 v^4 + 40680 q^7 v^5 + 1680 q^8 v^5 + 122384 q^6 v^6 + 10564 q^7 v^6 \cr
&+& 28 q^8 v^6 + 287760 q^5 v^7 + 46440 q^6 v^7 + 504 q^7 v^7 + 525096 q^4 v^8 \cr
&+& 149724 q^5 v^8 + 4272 q^6 v^8 + 718704 q^3 v^9 + 358456 q^4 v^9 + 22376 q^5 v^9 \cr
&+& 56 q^6 v^9 + 673920 q^2 v^{10} + 618792 q^3 v^{10} + 78864 q^4 v^{10} + 864 q^5 v^{10}\cr
&+& 331776 q v^{11} + 682128 q^2 v^{11} + 183864 q^3 v^{11} + 5976 q^4 v^{11} \cr
&+& 270432 q v^{12} + 225684 q^2 v^{12} + 20700 q^3 v^{12} + 96 q^4 v^{12} - 248832 v^{13}\cr
&-& 95256 q v^{13} + 10320 q^2 v^{13} + 264 q^3 v^{13} - 613008 v^{14} - 224916 q v^{14} \cr
&-& 11592 q^2 v^{14} - 108 q^3 v^{14} - 723600 v^{15} - 143536 q v^{15} - 3416 q^2 v^{15} \cr
&-& 8 q^3 v^{15} - 550377 v^{16} - 53844 q v^{16} - 444 q^2 v^{16} - 303288 v^{17}\cr
&-& 13512 q v^{17} - 24 q^2 v^{17} - 127636 v^{18} - 2284 q v^{18} - 41784 v^{19} \cr
&-& 240 q v^{19} - 10590 v^{20} - 12 q v^{20} - 2024 v^{21} - 276 v^{22} - 24 v^{23} - v^{24} \label{eq:K4}
\end{eqnarray}

\noindent
The 36-edge basis of Fig.~\ref{fig:kagomebases}c:
\begin{eqnarray}
& &q^{18} + 36 q^{17} v + 630 q^{16} v^2 + 7128 q^{15} v^3 + 12 q^{16} v^3 + 58503 q^{14} v^4 \cr
&+& 396 q^{15} v^4 + 370440 q^{13} v^5 + 6336 q^{14} v^5 + 1878670 q^{12} v^6 + 65322 q^{13} v^6 \cr
&+& 66 q^{14} v^6 + 7817940 q^{11} v^7 + 486456 q^{12} v^7 + 1980 q^{13} v^7 \cr
&+& 27122841 q^{10} v^8 + 2780142 q^{11} v^8 + 28710 q^{12} v^8 + 79228584 q^9 v^9 \cr
&+& 12642612 q^{10} v^9 + 267216 q^{11} v^9 + 220 q^12 v^9 + 195849552 q^8 v^{10} \cr
&+& 46787952 q^9 v^{10} + 1787790 q^{10} v^{10} + 6000 q^{11} v^{10} + 409851792 q^7 v^{11} \cr
&+& 142829784 q^8 v^{11} + 9120348 q^9 v^{11} + 78876 q^{10} v^{11} + 722497068 q^6 v^{12} \cr
&+& 361945104 q^7 v^{12} + 36695373 q^8 v^{12} + 662258 q^9 v^{12} + 585 q^{10} v^{12} \cr
&+& 1059589512 q^5 v^{13} + 760782804 q^6 v^{13} + 118640916 q^7 v^{13} \cr
&+& 3964788 q^8 v^{13} + 14424 q^9 v^{13} + 1261019664 q^4 v^{14} + 1313088660 q^5 v^{14} \cr
&+& 310356054 q^6 v^{14} + 17868048 q^7 v^{14} + 169878 q^8 v^{14} + 54 q^9 v^{14} \cr
&+& 1159847424 q^3 v^{15} + 1813868856 q^4 v^{15} + 652644588 q^5 v^{15} \cr
&+& 62170040 q^6 v^{15}+ 1258416 q^7 v^{15}+ 2472 q^8 v^{15} + 743510016 q^2 v^{16} \cr
&+& 1892524176 q^3 v^{16} + 1072172331 q^4 v^{16} + 167153268 q^5 v^{16} \cr
&+& 6467142 q^6 v^{16}+ 39798 q^7 v^{16} + 12 q^8 v^{16} + 251942400 q v^{17} \cr
&+& 1290209472 q^2 v^{17} + 1270157004 q^3 v^{17} + 334649052 q^4 v^{17} \cr
&+& 23635824 q^5 v^{17} + 347148 q^6 v^{17} + 528 q^7 v^{17} + 315207936 q v^{18} \cr
&+& 827146404 q^2 v^{18} + 431966778 q^3 v^{18} + 57655440 q^4 v^{18} \cr
&+& 1800882 q^5 v^{18} + 8596 q^6 v^{18} - 188956800 v^{19} - 193284144 q v^{19} \cr
&+& 116524980 q^2 v^{19} + 59677656 q^3 v^{19} +  4411884 q^4 v^{19} + 51288 q^5 v^{19} \cr
&+& 12 q^6 v^{19} - 678249936 v^{20} - 773787816 q v^{20} - 180556083 q^2 v^{20} \cr
&-& 11670198 q^3 v^{20} - 299721 q^4 v^{20} - 3066 q^5 v^{20} - 1195116768 v^{21} \cr
&-& 914327676 q v^{21} - 159944736 q^2 v^{21} - 8338004 q^3 v^{21} - 126372 q^4 v^{21} \cr
&-& 336 q^5 v^{21} - 1387555272 v^{22} - 681761826 q v^{22} - 75759996 q^2 v^{22} \cr
&-& 2362578 q^3 v^{22} - 17106 q^4 v^{22} - 12 q^5 v^{22} - 1197832536 v^{23} \cr
&-& 374476536 q v^{23} - 25228524 q^2 v^{23} - 441540 q^3 v^{23} - 1308 q^4 v^{23} \cr
&-& 819069489 v^{24} - 161325810 q v^{24} - 6339726 q^2 v^{24} - 59194 q^3 v^{24} \cr
&-& 48 q^4 v^{24} - 459630612 v^{25} - 56009760 q v^{25} - 1218540 q^2 v^{25} \cr
&-& 5664 q^3 v^{25} - 215826246 v^{26} - 15791340 q v^{26} - 175608 q^2 v^{26} \cr
&-& 360 q^3 v^{26} - 85595748 v^{27} - 3593976 q v^{27} - 17976 q^2 v^{27} \cr
&-& 12 q^3 v^{27} - 28735887 v^{28} - 648276 q v^{28} - 1167 q^2 v^{28} - 8138088 v^{29} \cr
&-& 89604 q v^{29} - 36 q^2 v^{29} - 1926908 v^{30} - 8946 q v^{30} - 375648 v^{31} \cr
&-& 576 q v^{31} - 58863 v^{32} - 18 q v^{32} - 7140 v^{33} - 630 v^{34} - 36 v^{35} - v^{36} \label{eq:K2x3}
\end{eqnarray}

\noindent
The 36-edge basis of Fig.~\ref{fig:kagomebases}d:
\begin{eqnarray}
& &q^{18} + 36 q^{17} v + 630 q^{16} v^2 + 7128 q^{15} v^3 + 12 q^{16} v^3 + 58506 q^{14} v^4 \cr
&+& 396 q^{15} v^4 + 370548 q^{13} v^5 + 6336 q^{14} v^5 + 1880512 q^{12} v^6 + 65334 q^{13} v^6 \cr
&+& 66 q^{14} v^6 + 7837704 q^{11} v^7 + 486876 q^{12} v^7 + 1980 q^{13} v^7 \cr
&+& 27272076 q^{10} v^8 + 2787144 q^{11} v^8 + 28728 q^{12} v^8 + 80068176 q^9 v^9 \cr
&+& 12716088 q^{10} v^9 + 267888 q^{11} v^9 + 220 q^{12} v^9 + 199480572 q^8 v^{10} \cr
&+& 47328210 q^9 v^{10} + 1799508 q^{10} v^{10} + 6018 q^{11} v^{10} + 422114976 q^7 v^{11}\cr
&+& 145763088 q^8 v^{11} + 9245904 q^9 v^{11} + 79656 q^{10} v^{11} + 754976880 q^6 v^{12} \cr
&+& 374020200 q^7 v^{12} + 37614312 q^8 v^{12} + 676604 q^9 v^{12} + 612 q^{10} v^{12} \cr
&+& 1126577808 q^5 v^{13} + 798846408 q^6 v^{13} + 123476376 q^7 v^{13} \cr
&+& 4118160 q^8 v^{13} + 15432 q^9 v^{13} + 1366323552 q^4 v^{14} + 1404553176 q^5 v^{14}\cr
&+& 329074176 q^6 v^{14} + 18944598 q^7 v^{14} + 186240 q^8 v^{14} + 90 q^9 v^{14} \cr
&+& 1280396160 q^3 v^{15} + 1977683904 q^4 v^{15} + 706012560 q^5 v^{15} \cr
&+& 67401848 q^6 v^{15} + 1413336 q^7 v^{15}+ 3528 q^8 v^{15} + 834582528 q^2 v^{16}\cr
&+& 2099656080 q^3 v^{16} + 1181501136 q^4 v^{16} + 184956942 q^5 v^{16} \cr
&+& 7412292 q^6 v^{16} + 53976 q^7 v^{16} + 36 q^8 v^{16} + 286654464 q v^{17} \cr
&+& 1451444832 q^2 v^{17} + 1419423912 q^3 v^{17} + 375629268 q^4 v^{17}\cr
&+& 27409380 q^5 v^{17} + 457176 q^6 v^{17} + 1152 q^7 v^{17} + 358691328 q v^{18} \cr
&+& 931971420 q^2 v^{18} + 487209582 q^3 v^{18} + 66732144 q^4 v^{18} \cr
&+& 2296374 q^5 v^{18} + 15436 q^6 v^{18} + 6 q^7 v^{18} - 214990848 v^{19} \cr
&-& 217611360 q v^{19} + 128909448 q^2 v^{19} + 67192668 q^3 v^{19} \cr
&+& 5321868 q^4 v^{19} + 80928 q^5 v^{19} + 144 q^6 v^{19} - 765345024 v^{20} \cr
&-& 871306632 q v^{20} - 208435194 q^2 v^{20} - 14587674 q^3 v^{20} \cr
&-& 406632 q^4 v^{20} - 4050 q^5 v^{20} - 1334035008 v^{21} - 1023060564 q v^{21} \cr
&-& 184223328 q^2 v^{21} - 10352660 q^3 v^{21} - 178764 q^4 v^{21} - 540 q^5 v^{21} \cr
&-& 1528264908 v^{22} - 755329716 q v^{22} - 87028104 q^2 v^{22} - 2969586 q^3 v^{22} \cr
&-& 25248 q^4 v^{22} - 18 q^5 v^{22} - 1299190104 v^{23} - 409417512 q v^{23} \cr
&-& 28747908 q^2 v^{23} - 554952 q^3 v^{23} - 1956 q^4 v^{23} - 873980685 v^{24}\cr
&-& 173604762 q v^{24} - 7119900 q^2 v^{24} - 72934 q^3 v^{24} - 72 q^4 v^{24} \cr
&-& 482643540 v^{25} - 59258256 q v^{25} - 1341372 q^2 v^{25} - 6684 q^3 v^{25} \cr
&-& 223371438 v^{26} - 16435896 q v^{26} - 188820 q^2 v^{26} - 396 q^3 v^{26} \cr
&-& 87528036 v^{27} - 3687672 q v^{27} - 18852 q^2 v^{27} - 12 q^3 v^{27} \cr
&-& 29116983 v^{28} - 657756 q v^{28} - 1194 q^2 v^{28} - 8194248 v^{29} \cr
&-& 90204 q v^{29} - 36 q^2 v^{29} - 1932752 v^{30} - 8964 q v^{30} - 376032 v^{31}\cr
&-& 576 q v^{31} - 58875 v^{32} - 18 q v^{32} - 7140 v^{33} - 630 v^{34} - 36 v^{35} - v^{36} \label{eq:K3x2}
\end{eqnarray}


\bigskip

\begin{thebibliography}{99}

\bibitem{Potts52} R.B.~Potts,
  Proc.~Camb.~Phil.~Soc.~{\bf 48}, 106--109 (1952).

\bibitem{Wu82} F.Y.\ Wu,
  Rev.\ Mod.\ Phys.\ {\bf 54}, 235 (1982).

\bibitem{Baxter_book} R.J.\ Baxter,
   {\em Exactly solved models in statistical mechanics},
  (Academic Press, London, 1982).

\bibitem{FK72} C.M.~Fortuin and P.W.~Kasteleyn,
  Physica {\bf 57}, 536--564 (1972).

\bibitem{Onsager44} L.\ Onsager,
  Phys.\ Rev.\ {\bf 65}, 117--149 (1944).

\bibitem{Baxter73} R.J.~Baxter,
  J.~Phys.~C {\bf 6}, L445--L448 (1973).

\bibitem{Baxter78} R.J.\ Baxter, H.N.V.\  Temperley and S.E.\ Ashley,
  Proc.\ R.\ Soc.\ Lond.\ A {\bf 358}, 535--559 (1978).

\bibitem{Wu10a} F.Y.\ Wu,
  Phys.\ Rev.\ E {\bf 81}, 061110 (2010).

\bibitem{JacSal06} J.L. Jacobsen and H. Saleur,
  Nucl.\ Phys.\ B {\bf 743}, 207--248 (2006).

\bibitem{Baxter82} R.J.\ Baxter,
  Proc.\ Roy.\ Soc.\ London Ser.\ A {\bf 383}, 43 (1982).

\bibitem{Wu79} F.Y.\ Wu,
  J.\ Phys.\ C {\bf 12}, 645 (1979).

\bibitem{Tsallis82} C.\ Tsallis,
  J.\ Phys.\ C {\bf 15}, L757 (1982).

\bibitem{KanoNaya53} K.\ Kano and S.\ Naya,
  Prog.\ Theor.\ Phys.\ {\bf 10}, 158 (1953).

\bibitem{Barry88} J.H.\ Barry, M.\ Khatun and T.\ Tanaka,
  Phys.\ Rev.\ B {\bf 37}, 5193 (1988).

\bibitem{JSS05} J.L. Jacobsen, J. Salas and A.D. Sokal,
  J.\ Stat.\ Phys.\ {\bf 119}, 1153--1281 (2005).

\bibitem{ZiffGu09} R.M.\ Ziff and H.\ Gu,
  Phys.\ Rev.\ E {\bf 79}, 020102(R) (2009).

%
\bibitem{ZiffSuding97} R.M.\ Ziff and P.N.\ Suding,
  J.\ Phys.\ A: Math.\ Gen.\ {\bf 30}, 5351 (1997).

\bibitem{Jensen97} I.\ Jensen, A.J.\ Guttmann and I.G.\ Enting,
  J.\ Phys.\ A: Math.\ Gen.\ {\bf 30}, 8067 (1997).

\bibitem{Monroe03} J.L.\ Monroe,
  Phys.\ Rev.\ E {\bf 67}, 017103 (2003).

\bibitem{Wu10b} C.\ Ding, Z.\ Fu, W.\ Guo and F.Y.\ Wu,
  Phys.\ Rev.\ E {\bf 81}, 061111 (2010).

\bibitem{Salas08} J.L.\ Jacobsen and J.\ Salas, unpublished (2008).

\bibitem{Saleur91} H.\ Saleur,
  Nucl.\ Phys.\ B {\bf 360}, 219 (1991).

\bibitem{Sokal05} A.D.\ Sokal, {\em The multivariate Tutte polynomial
    (alias Potts model) for graphs and matroids}, in B.S.\ Webb (ed.)
  {\em Surveys in combinatorics}, Lond.\ Math.\ Soc.\ Lect.\ Note
  Ser.\ {\bf 327}, 173 (2005).

\bibitem{Scullard11} C.R.\ Scullard,
  {\tt arXiv:1111.1061}.

\bibitem{ScullardZiff08} C.R.\ Scullard and R.M.\ Ziff,
  Phys.\ Rev.\ Lett. {\bf 100}, 185701 (2008).

\bibitem{ScullardZiff10} C.R.\ Scullard and R.M.\ Ziff,
  J.\ Stat.\ Mech. {\bf 2010}, P03021. 

\bibitem{Scullard11JSM} C.R.\ Scullard,
  J.\ Stat.\ Mech. {\bf 2011}, P09022.

\bibitem{Scullard12} C.R.\ Scullard, In preparation.

\bibitem{WuLin80} F.Y.~Wu and K.Y.~Lin,
  J.~Phys.~A {\bf 13}, 629--636 (1980).

\bibitem{Enting87} I.G.\ Enting,
  J.\ Phys.\ A: Math.\ Gen.\ {\bf 20}, L917 (1987).

\bibitem{Wu_priv} F.Y.\ Wu, private communication.

\bibitem{Maillard02} J.-M.\ Maillard,
  Chinese J.\ Phys.\ {\bf 40}, 327--378 (2002);
  {\tt arXiv:cond-mat/0205063v2}.

\bibitem{Scullard06} C.R.\ Scullard,
  Phys.\ Rev.\ E {\bf 73}, 016107 (2006).

\bibitem{Ziff06} R.M.\ Ziff,
  Phys.\ Rev.\ E {\bf 73}, 016134 (2006).

\bibitem{Wu06} F.Y.\ Wu,
  Phys.\ Rev.\ Lett.\ {\bf 96}, 090602 (2006).

\bibitem{Salas06} J.L.\ Jacobsen and J.\ Salas,
  J.\ Stat.\ Phys.\ {\bf 122}, 705--760 (2006).

\bibitem{Salas07} J.L.\ Jacobsen and J.\ Salas,
  Nucl.\ Phys.\ B {\bf 783}, 238--296 (2007).

\bibitem{Huse92} D.A.\ Huse and A.D.\ Rutenberg,
  Phys.\ Rev.\ B {\bf 45}, 7536--7539 (1992).

\bibitem{JacSal08} J.L. Jacobsen and H. Saleur,
  J.\ Stat.\ Phys.\ {\bf 132}, 707--719 (2008).

\bibitem{MooreNewman00} C.M.\ and M.E.J.\ Newman,
  J.\ Stat.\ Phys.\ {\bf 99}, 629--660 (2000).

\bibitem{cFSS1} H.W.J.\ Bl\"ote, J.L.\ Cardy and M.P.\ Nightingale,
  Phys.\ Rev.\ Lett.\ {\bf 56}, 742 (1986).

\bibitem{cFSS2} I.\ Affleck,
  Phys.\ Rev.\ Lett.\ {\bf 56}, 746 (1986).

\bibitem{Zamolo} A.B.\ Zamolodchikov,
  Sov.\ Phys.\ JETP Lett.\ {\bf 43}, 730 (1986)
  [Zh.\ Eksp.\ Teor.\ Fiz.\ {\bf 43}, 565 (1986)].

\bibitem{Cardy98} J.L.\ Jacobsen and J.L.\ Cardy,
  Nucl.\ Phys.\ B {\bf 515}, 701--742 (1998).

\bibitem{Blote82} H.W.J.\ Bl\"ote and M.P.\ Nightingale,
  Physica A {\bf 112}, 405--465 (1982).

\bibitem{SS01} J.\ Salas and A.D.\ Sokal,
  J.\ Stat.\ Phys.\ {\bf 104}, 609--699 (2001).

\bibitem{TL71} H.N.V.\ Temperley and E.H.\ Lieb,
  Proc.\ R.\ Soc.\ Lond.\ A  {\bf 322}, 251--280 (1971).  

\bibitem{Baek11} S.K.\ Baek, H.\ M\"akel\"a, P.\ Minnhagen and B.J.\ Kim,
  Phys.\ Rev.\ E {\bf 83}, 061104 (2011).

\bibitem{KSS08} R.\ Koteck\'y, J.\ Salas and A.D.\ Sokal,
  Phys.\ Rev.\ Lett.\ {\bf 101}, 030601 (2008).

\end{thebibliography}
\end{document}